\documentclass[review]{elsarticle}

\usepackage{url,graphicx,epsfig,graphics,subfigure,psfrag,amsmath,amssymb,epstopdf,enumerate,lineno,hyperref}
\modulolinenumbers[5]

\usepackage{diagbox}
\usepackage{xcolor}
\usepackage[normalem]{ulem}
\newcommand{\stkout}[1]{\ifmmode\text{\sout{\ensuremath{#1}}}\else\sout{#1}\fi}

\journal{Physica A}

\begin{document}

\begin{frontmatter}

\title{
Applications of Domain Adversarial Neural Network in phase transition of 3D Potts model}
%
\author[mymainaddress]{Xiangna Chen}

\author[mymainaddress,secondaryaddress]{Feiyi Liu\corref{mycorrespondingauthor}}
\cortext[mycorrespondingauthor]{Corresponding author}
\ead{fyliu@mails.ccnu.edu.cn}

\author[mymainaddress]{Weibing Deng\corref{mycorrespondingauthor1}}
\cortext[mycorrespondingauthor1]{Co-Corresponding author}
\ead{wdeng@mail.ccnu.edu.cn}


\author[shiyangaddress]{Shiyang Chen}
\author[Fourthaddress]{Jianmin Shen}

\author[secondaryaddress]{G\'abor Papp}
\author[mymainaddress]{Wei Li}
\author[mymainaddress]{Chunbin Yang}

\address[mymainaddress]{Key Laboratory of Quark and Lepton Physics (MOE) and Institute of Particle Physics, Central China Normal University, Wuhan 430079, China}

\address[secondaryaddress]{Institute for Physics, E{\"o}tv{\"o}s Lor\'and University\\1/A P\'azm\'any P. S\'et\'any, H-1117, Budapest, Hungary}

\address[shiyangaddress]{Department of Physics, Swansea university, SA2 8PP, Swansea, United Kingdom}

\address[Fourthaddress]{School of Engineering and Technology, Baoshan University,  Baoshan 678000, China}


\begin{abstract}

Machine learning techniques exhibit significant performance in discriminating different phases of matter and provide a new avenue for studying phase transitions.
We investigate the phase transitions of three dimensional $q$-state Potts model on  cubic lattice by using a transfer learning approach, Domain Adversarial Neural Network (DANN). 
With the unique neural network architecture, it could evaluate the high-temperature (disordered) and low-temperature (ordered) phases, and identify the first and second order phase transitions. 
Meanwhile, by training the DANN with a few labeled configurations, the critical points for $q=2,3,4$ and $5$ can be predicted with high accuracy, which are consistent with those  of the Monte Carlo simulations.
These findings would promote us to learn and explore the properties of phase transitions in high-dimensional systems.

\end{abstract}

\begin{keyword}
Machine learning \sep Domain adversarial neural network \sep  3D Potts model \sep  Phase transitions \sep Critical phenomena
\end{keyword}

\end{frontmatter}



\section{Introduction}
\label{intro}

Phase transition theory is one of the most interesting fields of condensed matter physics due to the existence of critical phenomena, and it has always played an important role in statistical physics
\cite{domb2000phase,stanley1971phase,plischke1994equilibrium,reichl2016modern,yeomans1992statistical,landau2013statistical}.
Excitingly, with the fast development of neural network (NN) algorithms and computing power, machine learning (ML) provides alternative solutions for traditional physical problems
\cite{carleo2019machine,zhang2018deep,torlai2016learning,zhang2017quantum,carleo2017solving}. For examples, the classification and regression in particle physics \cite{baldi2016jet,komiske2018learning,ma2022jet} and cosmology
\cite{brammer2008eazy,lanusse2018cmu,ravanbakhsh2016estimating}, the optimization of many-body simulations in many-body quantum matter
\cite{sharir2020deep,arsenault2017projected,huang2022provably,broecker2017machine}, the noise mitigation in optical fiber communication systems \cite{nevin2021machine,fan2020advancing,amari2019machine}, etc.
Especially, ML has also been applied to studying the phase transitions in statistical physics
\cite{carrasquilla2017machine,van2017learning,
zhang2019machine,tanaka2017detection,tomita2020machine,wang2016discovering}.

As a classic model of spin systems, the $q$-state Potts model is considered to be an extension of the Ising model~\cite{potts1952some}. Phase transitions of the Potts model have been widely studied \cite{wang1989antiferromagnetic,artun2020complete,susskind1979dynamics}, such as critical points, critical exponents, entropy, spontaneous symmetry breaking, etc. As a significant paradigm of studying critical phenomena,
the $q$-state Potts model has its own important characteristics:
the magical dependence of its critical properties (phase transition) on the parameter $q$, a remarkable behavior to study.

The two dimensional Potts model has been systematically studied through numerical methods \cite{potts1952some,wu1982potts,
baxter1973potts,den1979relation,
baxter2016exactly,2004Calculation,iino2019detecting,hu1989monte}, and also some ML methods \cite{li2018applications,
yau2022generalizability,
tan2021universal,tan2020comprehensive,giataganas2022neural}.
 The critical temperature of the two dimensional
$q$-state Potts model on square lattice is theoretically given by
$T_{c}=1/ \ln(1+\sqrt{q})$ ~\cite{baxter1973potts,wu1982potts}. And regarding the change of critical behavior for $q$, an unambiguous conclusion is
that the Potts model on square lattice exhibits a second-order phase transition for $q \le 4$ and a first-order phase transition for $q>4$ \cite{wu1982potts,baxter1973potts}.

For the three dimensional situation, 
various numerical methods have been attempted to  explore the phase transition properties of the Potts model \cite{blote1979first,gendiar2002latent,bazavov2008phase,
hartmann2005calculation,2022Tensor,fukugita1989correlation,lee1991three}. 
 As far as we know, for the three dimension $q$-state Potts model, the phase transition is considered to be second-order for $q = 2$, while first-order for $q \geq 3$ \cite{bazavov2008phase}.
However, it should be pointed out that, compared to the two dimensional Potts model, it is no such analogous expression in three dimension to indicate the dependence of the critical temperature $T_c$ on $q$.
Therefore, using ML in high-dimensional systems is a feasible approach, and some exploratory attempts have been made on three dimensional $q$-state Potts models~\cite{tan2020comprehensive,yau2022generalizability}.
For ML methods, most of them only require raw configurations for input in the study
 of phase transition, like most traditional methods,
and this kind of data do not rely on the identification of any order parameters. 
By sufficiently input data, abstract features containing dense information can be heuristically recognized by ML, for phase identification.

As the two categories of ML, supervised learning and unsupervised learning have shown good performance in several paradigmatic models \cite{carrasquilla2020machine,tanaka2017detection}.
Considering their respective limitations:
the ``prior"  nature of supervised learning (all data should be correctly and accurately labeled); the incompleteness of unsupervised learning, for the fact that
the logical connections or dynamic characteristics of data cannot be fully recognized without sufficient input data. 
Therefore, the ML methods in these two categories may not be optimal for high-dimensional systems, especially the ones critical behavior remains to be determined.

At present, a method of semi-supervised NN, Domain Adversarial Neural Network (DANN), is proposed as a sub-discipline of ML \cite{ajakan2014domain,ganin2016domain,farahani2021brief,huembeli2018identifying}.  DANN is a specific case of the transfer learning (TL) \cite{pan2009survey,zhu2005semi,weiss2016survey}. In this case, the data both in the source domain and target domain are employed to train the model.
Its basic idea is to select the same and transferable features representations in the source domain (labeled data) and the target domain (usually unlabeled data), and improve the performance of the NN in target domain by using the knowledge from source domain. The impressive predictive power of DANN in the study of critical behaviors, like determining critical point and exponents, has been proved in the applications of several two dimensional models, such as the Ising model\cite{huembeli2018identifying}, the Bose-Hubbard model\cite{ch2018unsupervised}, the Su-Schrieffer-Heeger model with disorder \cite{huembeli2018identifying},  the site percolation and direct percolation (DP) model \cite{shen2021transfer}.  Also we have tried DANN to study the two-dimensional $q$-state Potts model and obtained encouraging results \cite{chen2023study}.

In this paper, we apply DANN to study the $3$D Potts model,
whose phase transition properties were the subjects of long-term debate \cite{2014Phase}.
Our results indicate that DANN is suitable for phase transition classification and critical point prediction in high-dimensional system (3D). Moreover, with its specific NN architecture, relevant information of phase transition types can be obtained for the three-dimensional Potts model, namely whether it is the first-order or second-order phase transition.

The paper is organized as follows.
In Section \ref{model}, we introduce the  $q$-state Potts  model and focus on three-dimensional situation. The methodology of applying the DANN method to the Potts model is demonstrated in Section \ref{method}. In Section \ref{sec:Results}, we analyze the results of the MC simulation and DANN.  Finally, we conclude in the last Section.

\section{Model}
 \label{model}
\subsection{ The q-state Potts model}
  \label{potts}

The three dimensional Potts model on a cubic lattice, which exhibits some differences in the form of the Hamiltonian $H$, can be represented as \cite{potts1952some,wu1982potts,chatelain20023d}:

\begin{equation}
\beta \mathcal{H}=-J \beta \sum _{\langle i,j \rangle}  \delta_{\sigma_i,\sigma_j}, \quad \sigma_i,\sigma_j\in \{0, 1,..., q-1\},
\end{equation}
where $\sigma_i$ is the spin  value at site $i$.
  $\langle i,j \rangle$ means the nearest neighbor pairs on the cubic lattice and
  the sum  is running on a cubic lattice $L \times L \times L$ with periodic boundary conditions.
 $\delta$ is the Kronecker delta.
 $\beta$ is the inverse temperature.
 $J$ is defined as the interaction constant and is set to be $1$ and temperature $T$ is measured in units of $J$. 
In this work, we focus on the critical behavior of this three dimensional Potts model with different $q$ state.
It is worth noting that the case of three dimension 
contains a larger amount of information than two dimension, and has more complex internal connections 
, which is a huge challenge for ML methods.

\subsection{Monte Carlo  Glauber Algorithm for Potts Model}\label{Glauberalgorithm}

The algorithms of generating the configuration samples of spin model include the Metropolis algorithm
\cite{mariz1990comparative}, the Glauber algorithm
\cite{meyer2000computational}, the Swendsen-Wang algorithm
\cite{swendsen1987nonuniversal}, the Wolff algorithm
\cite{wolff1989collective}, etc. 
Following our previous work \cite{chen2023study}, we still use the Glauber algorithm here.

For the three-dimensional Potts model with the periodic boundary conditions, the flip probability (transition probability) of a spin value $\sigma_i$ changing in next MC step is denoted by $p$.
The configuration gradually evolves according to the following spin-flipping mechanism: 
\begin{equation}
  \sigma_i=\left\{
\begin{array}{lll}
\sigma_i^{old}     &      & \text{if} \quad r>p, \quad \text{not accept flip}, \\
\sigma_i^{new}     &      & \text{if}  \quad r\leq p, \quad \text{accept flip}, 
\end{array} \right.
\end{equation}
where $r$ is a random number in $(0, 1)$,
while  $\sigma_i^{old}$ and $\sigma_i^{new}$ are the spin values of site $i$ before and after flipping,  which are randomly chosen from $\{0, 1,..., q-1\}$ and any of $q-1$ possible new values.

Based on the Glauber algorithm \cite{meyer2000computational},
the flip probability $p$ is defined as follows \cite{henkel2008non,glauber1963time,mariz1990comparative} :

\begin{equation}
p=\frac{1}{1+exp(\Delta \epsilon *\beta)}, \quad  \beta=\frac{1}{ T},
\end{equation}
where  $\Delta \epsilon =\epsilon^{new}-\epsilon^{old}$
indicates the energy difference of site $i$ before and after the flip. 
The energy $ \epsilon$ of site $i$ can be calculated by considering the sum of the interaction energies of that spin with its nearest neighbours:
\begin{align}
 \epsilon =  -(\delta_{\sigma_{xyz},\sigma_{(x-1)yz}}& +\delta_{\sigma_{xyz},\sigma_{(x+1)yz}} +\delta_{\sigma_{xyz},\sigma_{x(y+1)z}}
\notag
\\& +\delta_{\sigma_{xyz},\sigma_{x(y-1)z}}+\delta_{\sigma_{xyz},\sigma_{xy(z+1)}}+\delta_{\sigma_{xyz},\sigma_{xy(z-1)}}).
 \label{eq:energy}
\end{align}
here $x, y, z$ are the coordinates of site $i$, that is, $\sigma_{xyz}=\sigma_{i}$.

\begin{figure*}[htbp]
\centering
\subfigure[$q=5,T=0.05$]{
\label{Fig.configq3.1}
\includegraphics[width=0.45\textwidth]{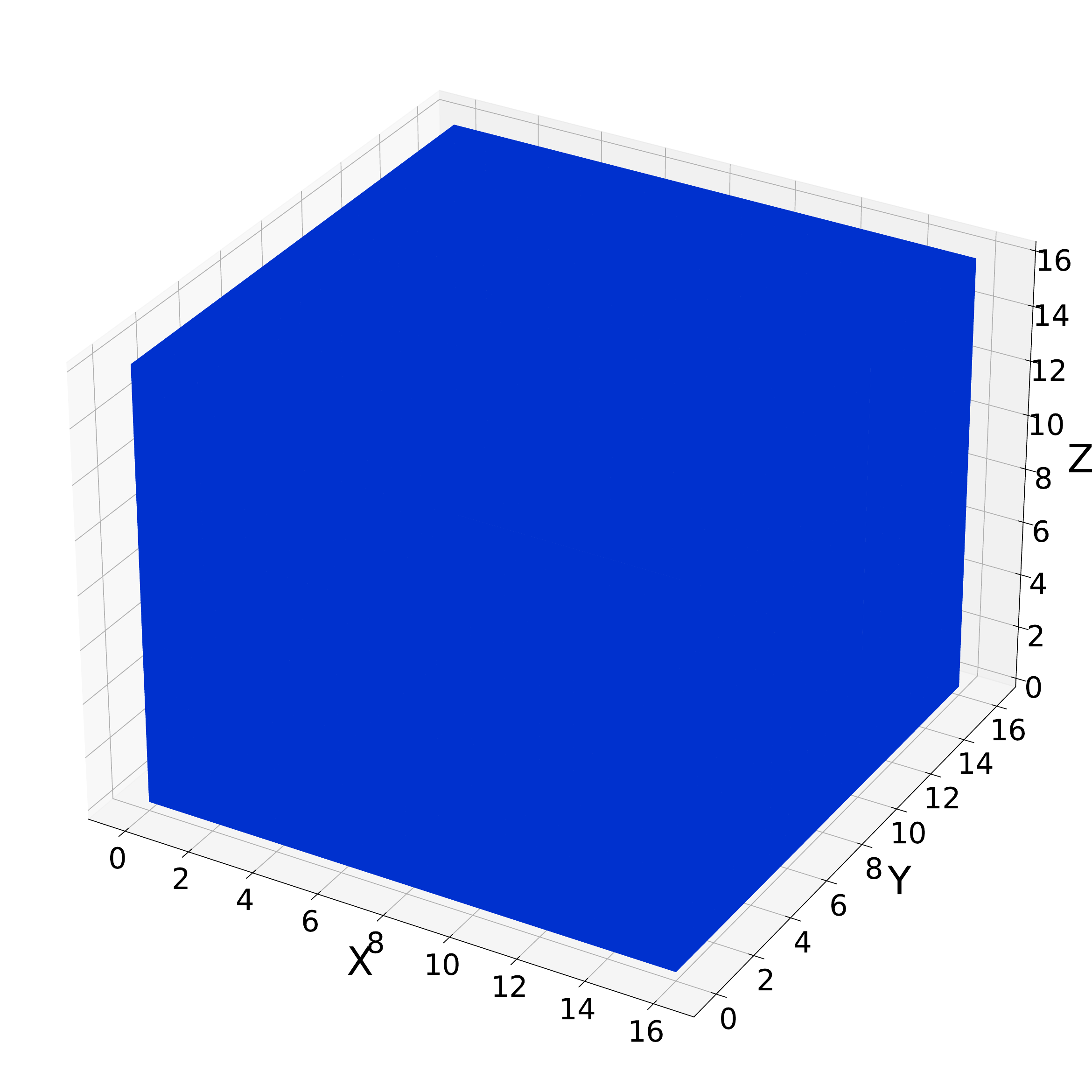}}
\subfigure[$q=5,T=1.45$]{
\label{Fig.configq3.3}
\includegraphics[width=0.45\textwidth]{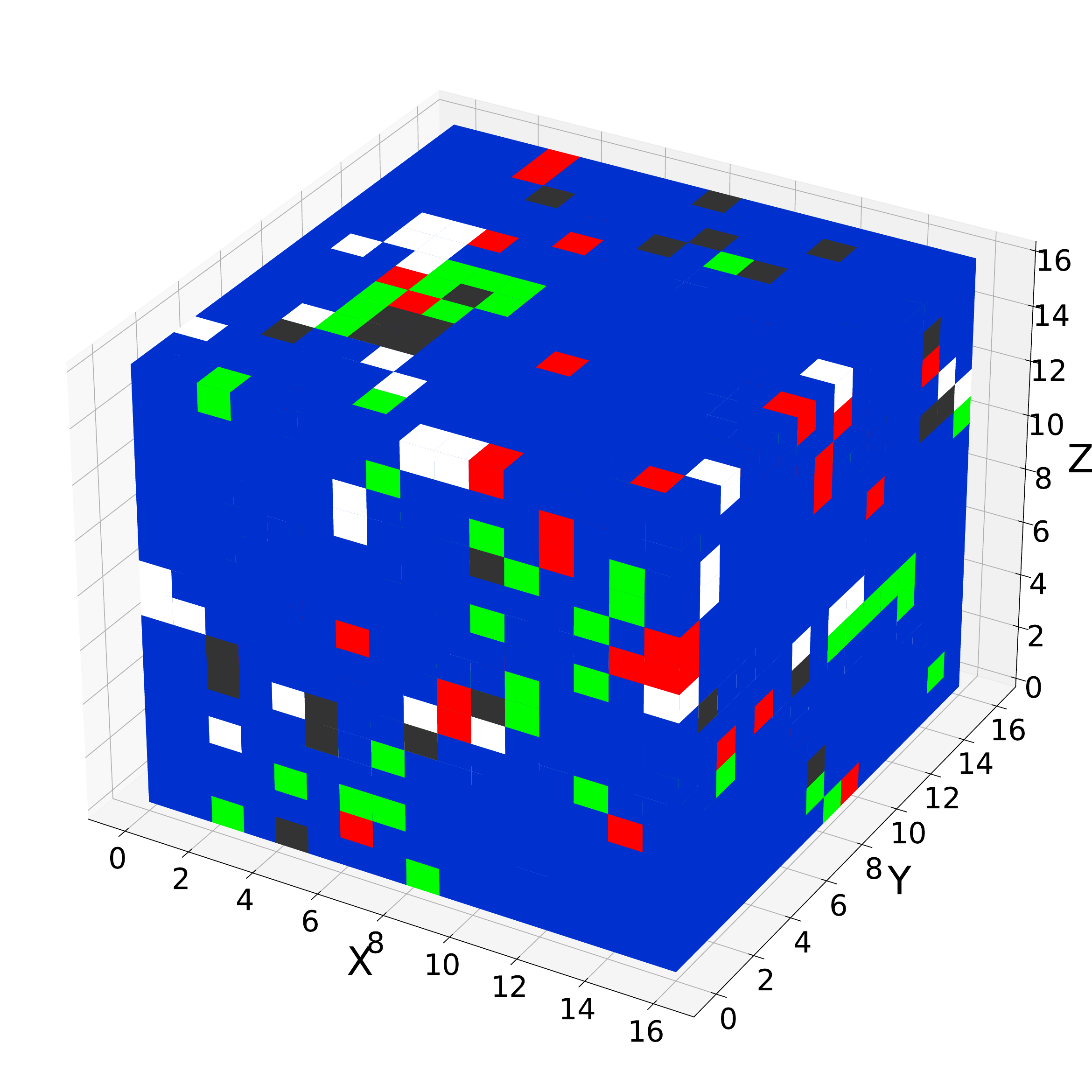}}
\subfigure[$q=5,T=3.0$]{
\label{Fig.configq3.3}
\includegraphics[width=0.45\textwidth]{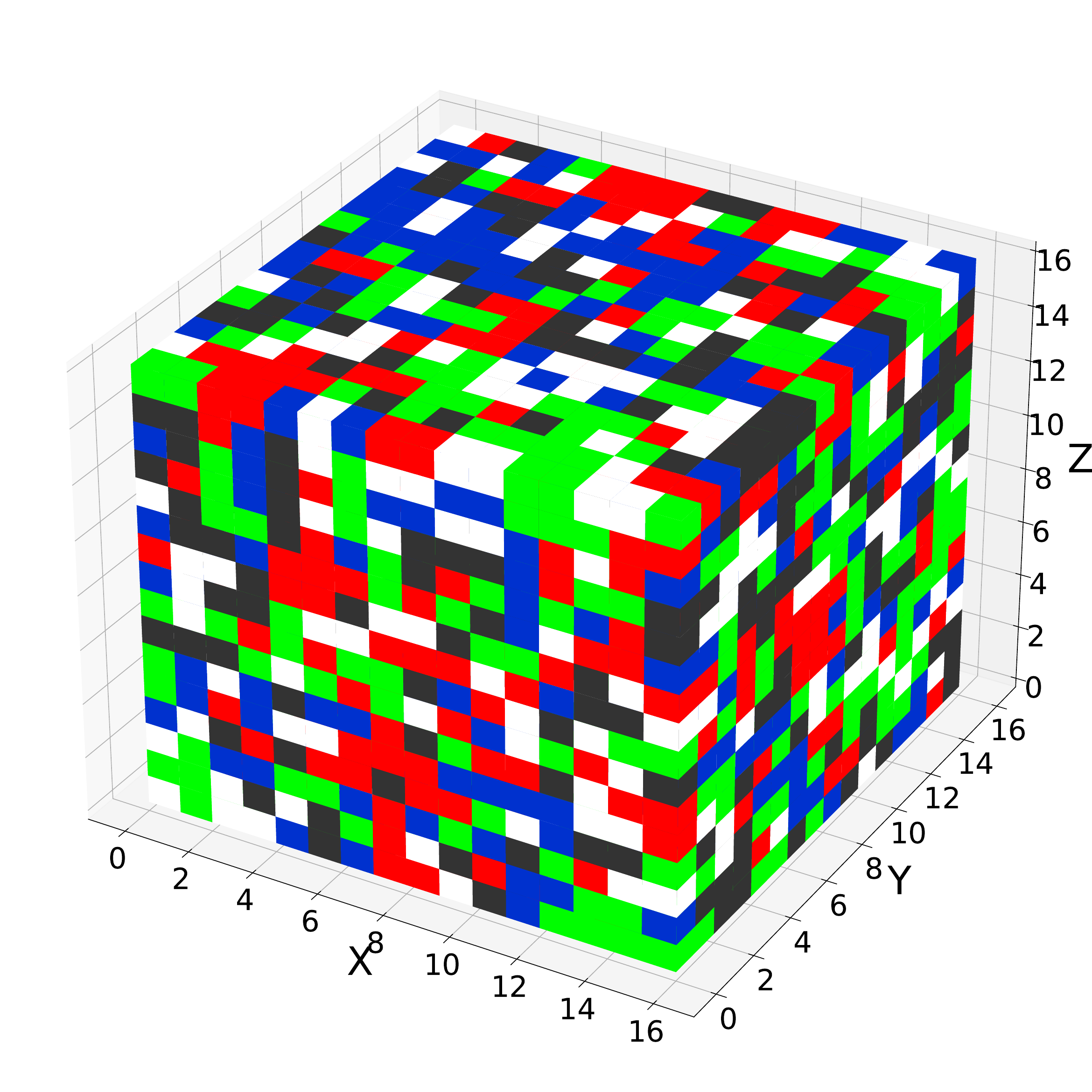}}
\\

\caption{The instantaneous snapshots of the $5$-state Potts model generated by MC simulation with $L = 16$ for (a) $T=0.05 < T_c$; (b) $T=1.45 \approx T_c$; (c) $T=3.0 > T_c$. The \{$X, Y, Z$\} is coordinates.
The color (red, blue,  green, black and white) of the unit lattice represents the spin value for each site $i$, i.e., $\sigma_i=0, 1, 2, 3$ and $4$, respectively. }
\label{fig:config}
\end{figure*}

Using the Glauber algorithm for updating spin's  configuration, we can generate the spin configurations with different values of $T$. Fig.~\ref{fig:config} shows
a few typical spin configurations of three-dimensional $5$-state Potts model with lattice size $L=16$, illustrating the changes of phase with respect to different temperatures from left to right.
For $T=0.05<T_c$, one can observe an ordered state (the full lattice is occupied by blue).
At $T=1.45$ close to $T_c$,
it can be observed that the system has become disordered
(blue dominates but other colors have emerged).
When $T=3.0$ is large enough ($>>T_c$), the system stays in a completely disordered state (colors of small clusters are messy and random occupied the whole cube).
This obvious feature allows DANN to explore the mysteries of the Potts model through phase diagrams with different temperatures.

The data of raw configrations generated from Monte Carlo (MC) simulations above are used as the input of DANN.  
These configurations are stored for  $96$ temperature values with $T \in  [0.05, 4]$, for $q = 2,3,4$ and $5$ in each lattice size.
Considering the impact arising from the finite size \cite{barber1983finite,fisher1972scaling,
privman1990finite}, here we use different lattice sizes $L = 8, 12,16, 24$ and $32$ respectively, to generate the configurations.  
For a given temperature $T$, the total number of configurations $n=1000$ are saved as samples.

\section{Method}
\label{method}
\subsection{The Domain Adversarial Neural Network (DANN) method}\label{dann}

The three-dimensional DANN adopts the idea from two-dimensional DANN in Ref.~\cite{shen2021transfer}, which also led to the successful identification of the nature of two-dimensional models.
Nevertheless, the structure of DANN is different for the input data in three dimension, and it should be improved to get accurate results. 
In this section, we will explain the details of the three-dimensional DANN in our study, as well as the labeling rules for domains.

\begin{figure}[htbp]
\centering
\includegraphics[width=1\textwidth]{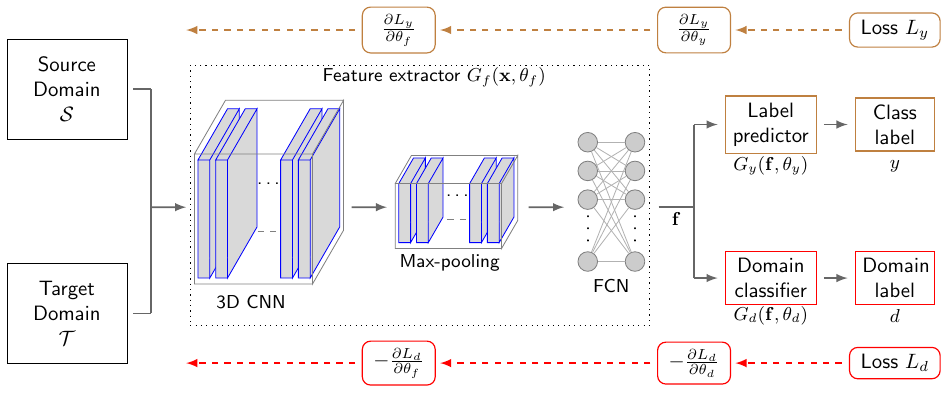}
\caption{The overall operating framework of DANN. The source domain $\mathcal{S}$ represents labeled data, the target domain $\mathcal{T}$ represents unlabeled data.
The neural network is composed of feature extractor $G_f(\mathbf{x},\theta _f)$, label predictor $G_y (\mathbf{f}, \theta _y)$ and  domain classifier $G_d (\mathbf{f}, \theta _d)$,
where $\mathbf{x}$ refers to input data, while $\theta _f$,  $\theta_y$, $\theta _d$ represent the network optimization parameters. $L$ stands for loss.}

\label{fig:DANNtk}
\end{figure}

Fig.~\ref{fig:DANNtk} shows the intuitive schematic structure of DANN, which is generally decomposed into three parts: feature extractor $G_f$, label predictor $G_y$ and domain classifier $G_d$. ( $G_f$ generates the features for both label predictor and domain classifier.)
The input data of DANN representing three-dimensional spin configurations are assigned to the source domain and target domain. We will introduce the labeling of data in detail later in section \ref{dataset}.
The source domain is written as $\mathcal{S}=\{(x_s, y_s)\}$ with data $x_s$ and label $y_s$, and the target domain is represented as $\mathcal{T}=\{ x_t \}$ with unlabeled data $x_t$.
Therefore, the given input variable $\mathbf{x}$ in $G_f(\mathbf{x},\theta _f)$ can be defined as:
$\mathbf{x}=\{x_s\}\cup \{x_t\}$.

About the $G_f(\mathbf{x},\theta _f)$:
The main role of $G_f(\mathbf{x},\theta _f)$ is to produce a high-dimensional feature vector $\mathbf{f}= G_f(\mathbf{x}, \theta _f)$ with parameter $\theta _f$ through feeding the input data $\mathcal{S}\cup\mathcal{T}$.
The structure of $G_f$ 
is mainly composed of 
a three-dimensional convolutional neural network (CNN) connected with a fully connected network (FCN) layer, as shown Fig.~\ref{fig:DANNtk}.
The input $x\in\{x_s\}\cup\{x_t\}$ of the NN are size of $L\times L\times L$ as three-dimensional images, which are convoluted into the CNN of $16$ filters by convolution operation forming the feature maps. 
The max-pooling layer is used here to reduce the feature maps by $1/2$. 
Later on, the feature maps are formatted by a flatten layer and output $[0,1]$ through a sigmoid activation.
Next, they are transferred to the NN: the fully connected layer with $50$ neurons.
After that, we additionally apply batch normalization (BN)
and the Relu activation 
to prevent overfitting and speed up the training process.
Finally, these feature vectors (latent variables) $\mathbf{f}$ are got through the entire $G_f(\mathbf{x},\theta _f)$, namely, $\mathbf{f}=G_f(\mathbf{x},\theta _f)$.

Subsequently, the latent variables $\mathbf{f}$ are fed into the label predictor $G_y (\mathbf{f}, \theta _y)$ and domain classifier $G_d (\mathbf{f}, \theta _d)$. 
For the specific structures, the label predictor has a dense network of 2 neurons with the softmax activation function, and the domain classifier contains only 1 neuron with hard sigmoid activation function. In the both structures, a BN is applied to make the process stable and faster, before the data flow passes through the activation function.
The output of $G_y (\mathbf{f}, \theta _y)$ is expressed as a vector $(P_0, P_1)$, namely, the predictive probabilities $P$ of the input configurations belonging to label ``0"  and label ``1", respectively. 
The domain classifier $G_d (\mathbf{f}, \theta _d)$ outputs the vector $d$ by judging whether the feature vector $\mathbf{f}$ is from the source domain $\mathcal{S}$ or the target domain $\mathcal{T}$, if from the former, outputs $d=0$, otherwise $d=1$.

The process incorporates the idea that the NN is trained in such a way that the feature representations of the two domains become invariant, rendering the domain classifier $G_d$ unable to differentiate between them (domains), and predict the corresponding labels for unlabeled data $x_t$ in target domain. This implies that we need to minimize the loss $L_{y}$ of the label predictor $G_y (\mathbf{f}, \theta _y)$ and  the loss $L_{d}$ of the domain classifier $G_d (\mathbf{f}, \theta _d)$.

To train this network, the loss function is defined as \cite{shen2021transfer,huembeli2018identifying}
:
\begin{equation}
L(\theta _f,\theta _y,\theta _d)=L_y(\theta _f,\theta _y)-L_d(\theta _f,\theta _d).
  \label{loss_Lf}
\end{equation}
where the the saddle point $\hat{\theta}_f$, $\hat{\theta}_y$ and $\hat{\theta}_d$:
\begin{align}
\hat{\theta}_f,\hat{\theta}_y =\mathop{\arg\min}\limits_{\theta _f,\theta _y} L(\theta _f,\theta _y, \hat{\theta}_d),\\
\hat{\theta}_d =\mathop{\arg\max}\limits_{\theta _d} L(\hat{\theta}_f,\hat{\theta}_y, \theta_d).	
\label{loss_argmin}
\end{align}
the optimization process involves a minimization with respect to parameters  $\theta_f$,  $\theta_y$, as well as a maximization with respect to the $\theta_d$.
The update rules are as follows (learning rate $\mu = 0.0001$):
\begin{align}
 \label{loss_f} &\theta_f \quad\leftarrow \quad \theta_f-\mu\left(\frac{\partial L_y}{\partial\theta_f}- \frac{\partial L_d}{\partial\theta_f}\right),\\ 
 \label{loss_y} &\theta_y \quad\leftarrow \quad \theta_y- \mu\left( \frac{\partial L_y}{\partial\theta_y}\right),\\ 
 \label{loss_d} &\theta_d \quad \leftarrow \quad  \theta_d- \mu \left( \frac{\partial L_d}{\partial\theta_d}\right).
 \end{align}
The parameters of these three parts are optimized to achieve our goals. This optimization process has an adversarial nature:
DANN learns parameter $\theta _f$   by minimizing loss $L_y$ and maximizing the loss $L_d$, respectively, parameter $\theta _y$ by minimizing loss $L_y$, parameter $\theta _d$ by minimizing the loss $L_d$; But at the same time, $\theta _f$ is determined by optimizing $ L(\theta _f,\theta _y,\theta _d)$.
Our study is based on the software library, TensorFlow-GPU $1.14$ and Python 3.8 on NVIDIA GeForce RTX 3080 Ti platform 
with 16GB memory and 234GB storage space.

\subsection{Data sets of models}
\label{dataset}

To implement the classification of the two phases in the Potts model, one needs to prepare the input datas: three-dimensional configurations, especially to label the configurations of source domain.
Regarding the initial definition of the source domain, to minimize potential errors 
(human intervention),
we label the two phases based on two temperature ranges far from the critical regime of phase transition.
Specifically, we assign a label of ``0'' to the ordered phase for  $T\ll T_c$, and a label of ``1'' to the disordered phase for $T\gg T_c$, as shown in Fig.~\ref{fig:config}.
Therefore, for $q=2$ ,
we set the samples in $T\in  [0.05, 0.1] \cup [3.9, 4]$ as the initial source domain, where the former has a label ``0'' and the latter has a label ``1''.
The rest is the target domain.

After the training process with epoch$=1000$, the DANN can predict the binary  classification of each configuration sample in the target domain at each value of $T$. By averaging the output separately, we can get
the average probability that a spin configuration belongs to the phase ``0'' (ordered phase) and ``1'' (disordered phase) as $P_0$ and $P_1$ respectively.
The critical point $T_c$ is defined as the temperature $T$ at $P_0=P_1=50\%$.

\subsection{The optimal source domain of DANN}\label{optimal}

To ensure that the DANN can capture the similarity between the two domains, one needs to appropriately expand the source domain support defined in the previous section, which means that the optimal source domain is filtered to obtain more accurate results.
Our iterative method based on interval expansion starting from $T\in[0.05, l^{(0)}] \cup [r^{(0)},4]$ can be defined as:
\cite{shen2021transfer}:
\begin{equation}
l^{(i+1)}=\frac{l^{(i)}+T_{c,q}^{(i)}}{2},\quad 
r^{(i+1)}=\frac{r^{(i)}+T_{c,q}^{(i)}}{2},
\label{equ.lr}
\end{equation}
where $l^{(0)}=0.1$ and $r^{(0)}=3.9$,
$i$ represents the $i$-th expansion, and $T\in[0.05, l^{(i)}] \cup [r^{(i)},4]$ is  the i-th source domain interval. $T_{c,q}^{(i)}$ is the critical value estimated by DANN on the $i$-th source domain interval.
Meanwhile, for each iteration, one needs to follow the following rules: at least 99\% of the output in the source domain is classified as category ``0'' or ``1''.
If the condition is met, the iteration  continues;
if not, the boundary value ($l$ or $r$) is
adjusted towards the  initial  value, e.g., $l^{(i+1),1} \to (l^{(i+1),0})/2$.
The process is halted when the expansion of the source domain is no longer feasible.
A concrete application can be observed in Fig.\;\ref{fig:q2Tc}(b). 

\section{Results}\label{sec:Results}

\subsection{MC results }
\label{MC}

\begin{figure}[htbp]
\centering
\subfigure[]{
\label{Magq41}
\includegraphics[height=0.43\textwidth]{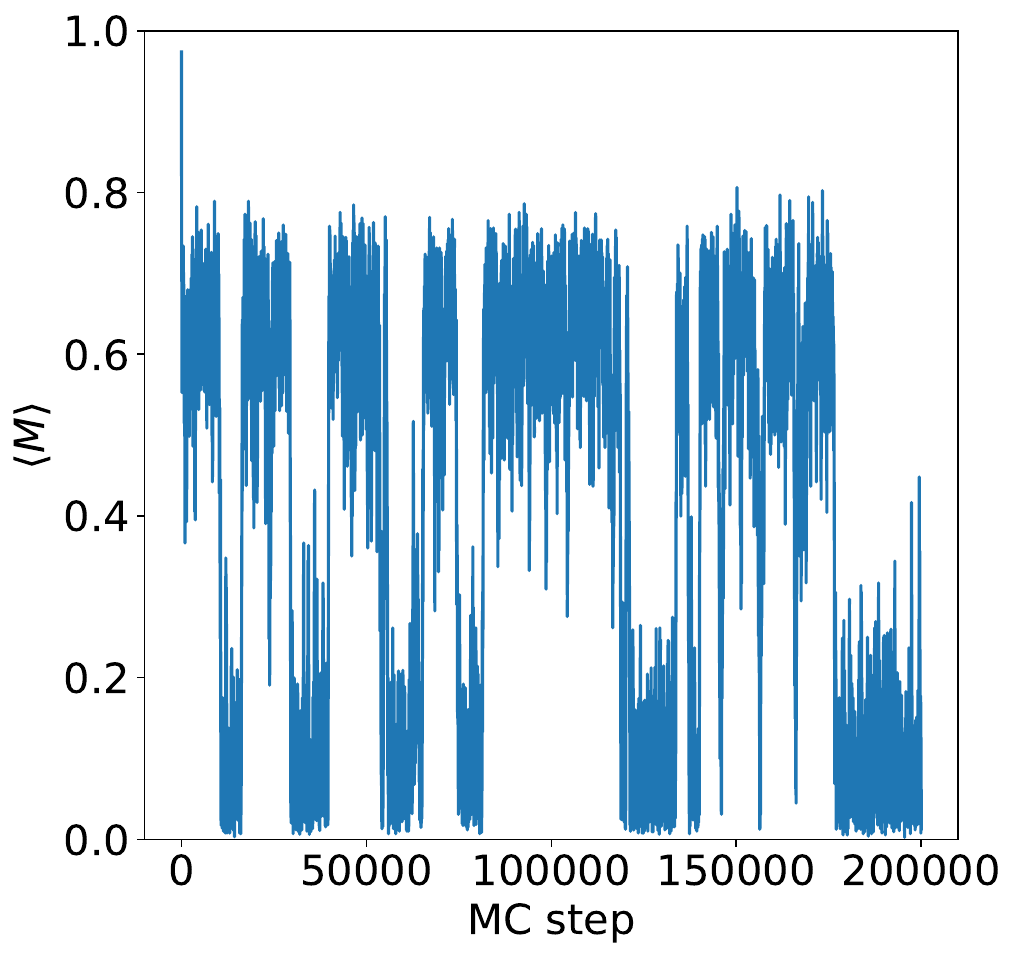}}
\subfigure[]{
\label{Eneq42}
\includegraphics[height=0.43\textwidth]{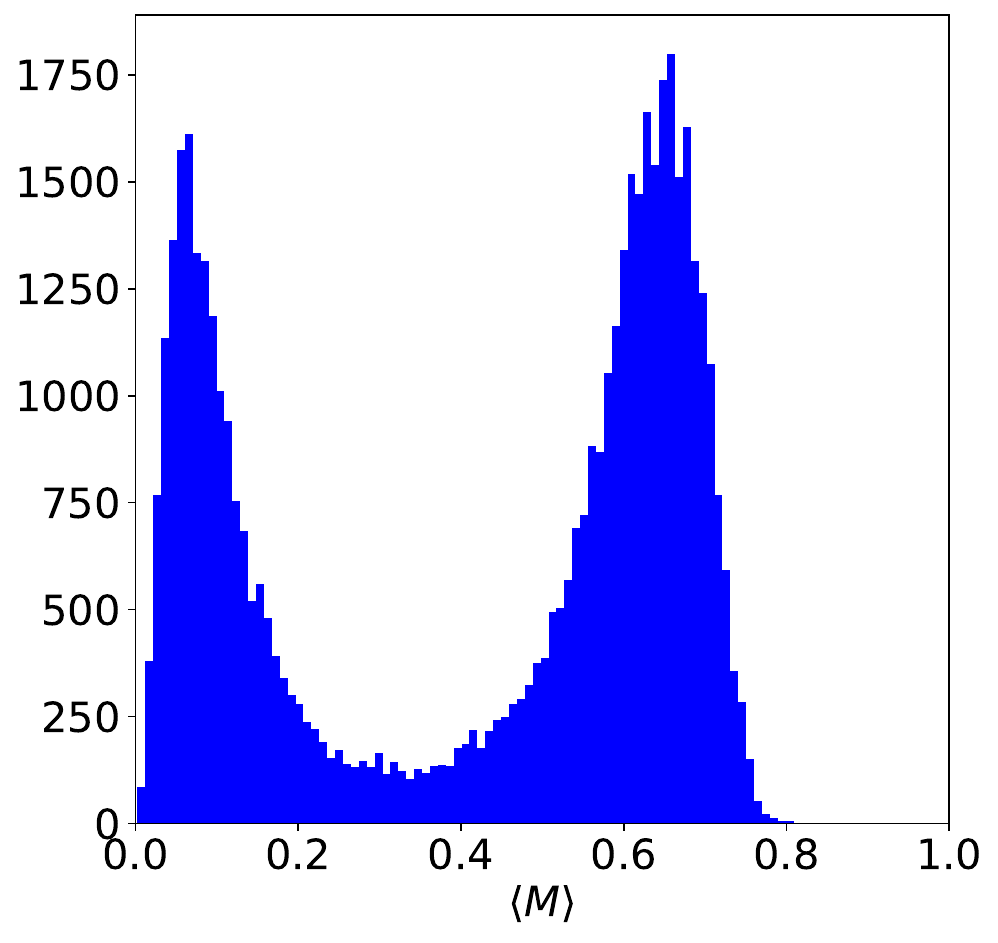}}
\caption{The  distribution of magnetization $\langle M \rangle$ for 3D $4$-states Potts model with $L=12$ at $T=1.592$. (a) $\langle M \rangle$ as a function of MC step. The total MC step is set to 200000. (b) The histogram of $\langle M \rangle$.}
\label{fig:Mq412}
\end{figure}

\begin{figure}[htbp]
\centering
\subfigure[]{
\label{Magq43}
\includegraphics[height=0.43\textwidth]{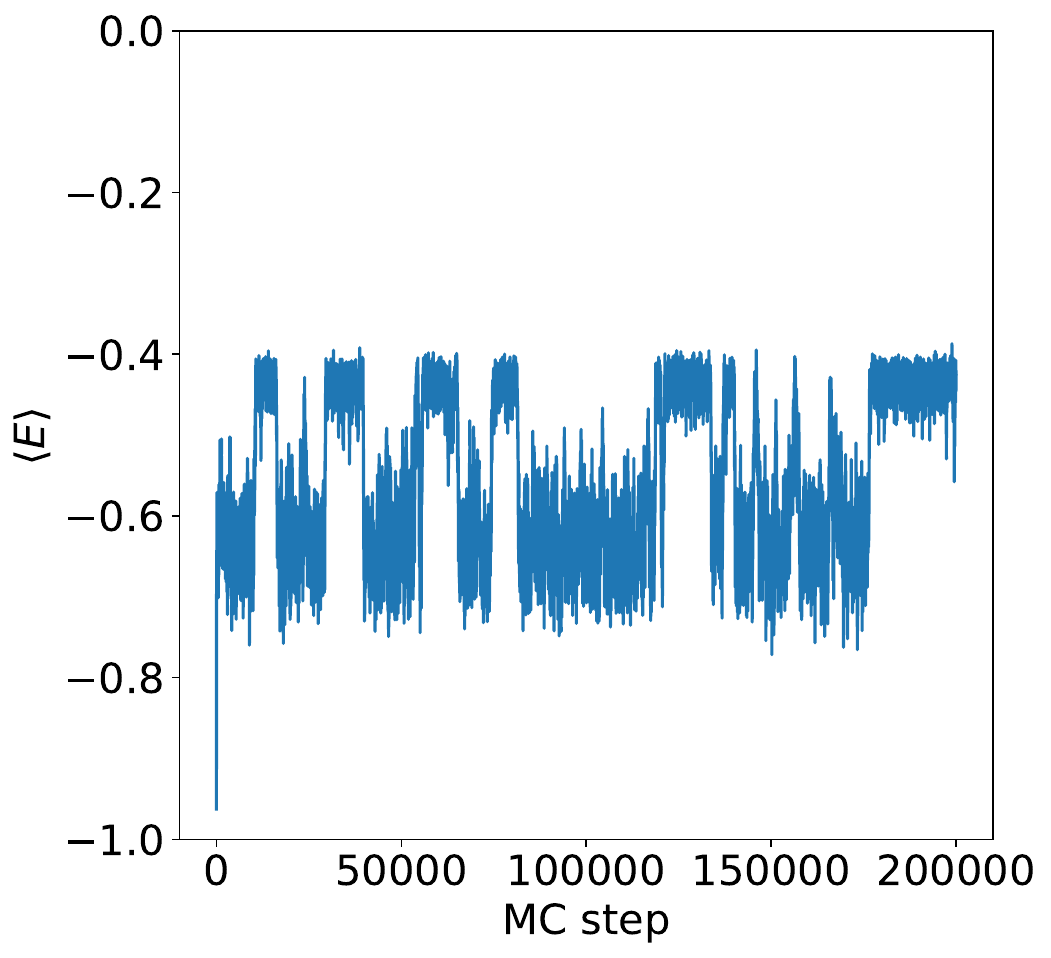}}
\subfigure[]{
\label{Eneq44}
\includegraphics[height=0.43\textwidth]{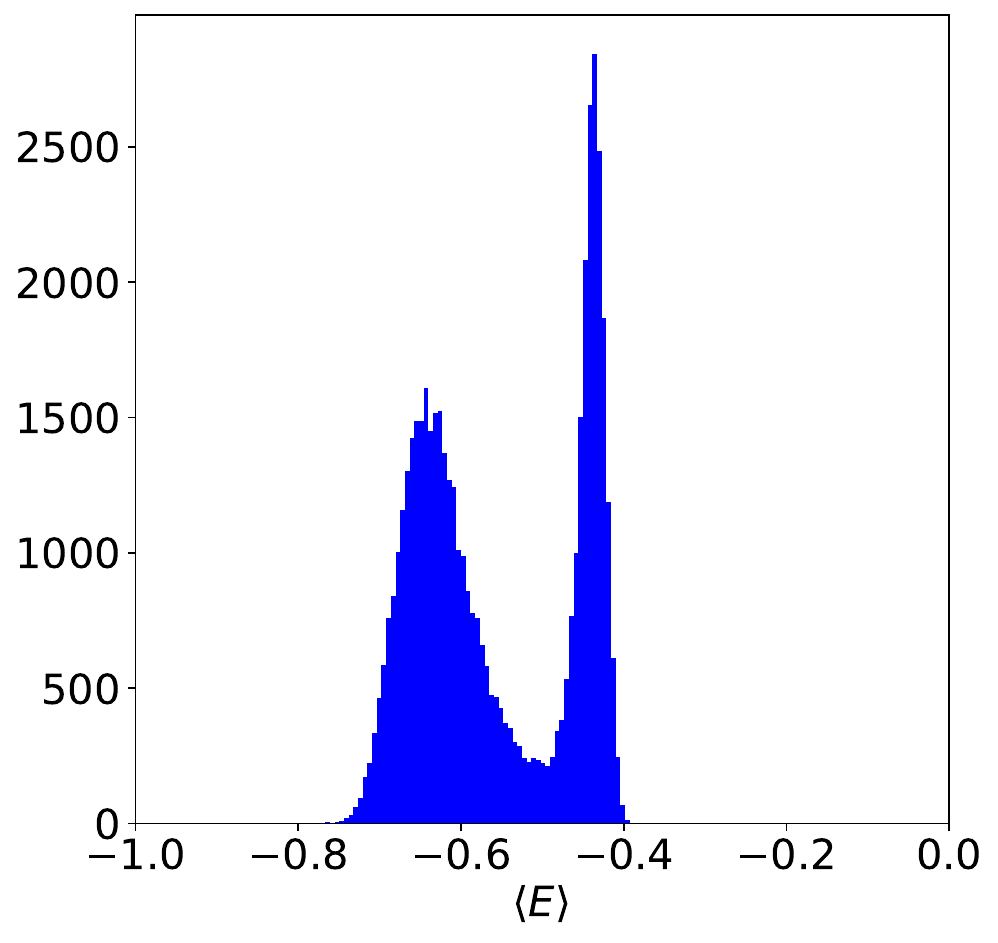}}
\caption{The  distribution of energy density $\langle E \rangle$ for 3D $4$-states Potts model with $L=12$ at $T=1.592$. (a) $\langle E \rangle$ as a function of MC step. The total MC step is set to 200000.(b) The histogram of $\langle E \rangle$.}
\label{fig:Mq412_E}
\end{figure}

To make sure the validity of the Glauber algorithm in generating spin configurations (for further exploration of the phase transition characteristics using machine learning), we firstly present the results of MC simulations of three dimensional $q$-state Potts model based on some relevant observables.
The observables we calculated here are the magnetization (per site) $\left| \langle M\rangle \right|$, energy $\langle E \rangle$,
where the order parameter $\langle M\rangle$ can be defined as 
\cite{henkel2008non,chatelain2005monte,fernandes2006alternative}:

\begin{equation}
\langle M \rangle=\frac{\frac{q}{{L^d}} \max(n_1,n_2,...,n_q)-1}{q-1},
\label{M}
\end{equation}
where $d=3$ and $L^d=L \times L \times L$ is the total number of sites.  And $n_q$ is the number of $q$-th spin value in each configuration.


A given configuration of spins in the cubic lattice $L \times L \times L$ has the energy per site:
\begin{equation}
\langle E\rangle= \frac{1}{L^d}  \sum _{i=1}^{L^d}  \varepsilon . 
\end{equation}
where $ \varepsilon $ is the energy defined in Eq.~(\ref{eq:energy}), and $\langle E \rangle$  is is calculated on average for $L^d$ sites.
In Fig.~\ref{Magq41}, we show the magnetization $\langle M \rangle$ 
of 3D $4$-states Potts model as a function of MC step, at $T=1.592$ close to $T_c$. 
Obvious, it jumps between two value up and down with the MC steps. 
Further, the histogram in Fig~\ref{Eneq42} shows a two peak structure which indicates a first order phase transition~\cite{tan2020comprehensive}.
A similar test of the energy density  $\langle E \rangle$ shown in Fig~\ref{fig:Mq412_E} is also an evidence to support this inference with the double-peak structure histogram.

In this section, we demonstrated the MC simulation and found that it is consistent with theoretical predictions regarding the types of phase transition, which indicates that the MC data is reliable.
Next, instead of relying on the internal information of the model or a certain order parameter, we only consider the raw configurations and use the DANN to learn the phase transition characteristics of the 3D Potts model.

\subsection{The DANN results of q=2}
\label{q2_result}

\begin{figure}[htbp]
\centering
\subfigure[]{
\label{Fig.q2.1}
\includegraphics[height=0.3\textwidth]{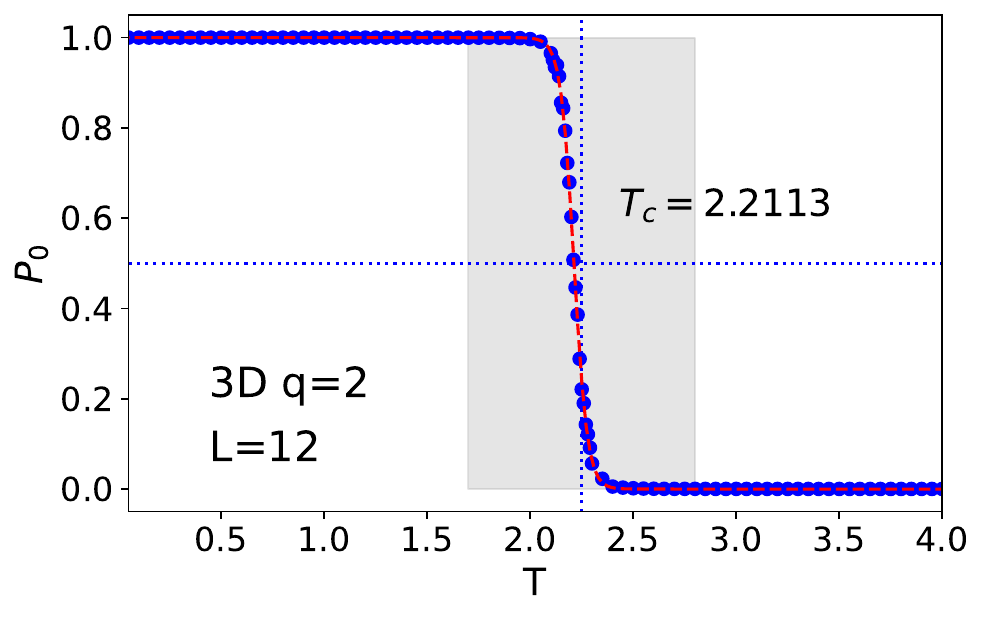}}
\subfigure[]{
\label{Fig.q2.2}
\includegraphics[height=0.3\textwidth]{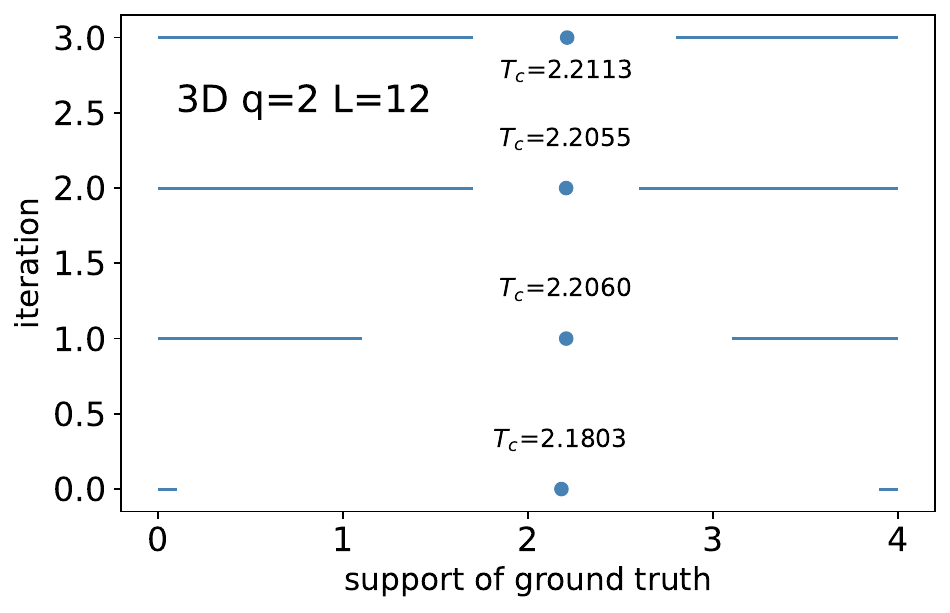}}
\caption{(a) Average probability $P_0$ of being in phase ``0'' for 3D $2$-state Potts mode with $L = 12$, as a function of temperature $T$. (b) The evolution process of the optimal domain and the corresponding detected critical temperatures.
The $x$-axis represents $T$, ranging from $0.05$ to $4$, while the $y$-axis denotes the $i$-th iteration.
The blue dots represent the estimated critical temperature $T_c$ by DANN, obtained during the $i$-th iteration on the domain interval.}
\label{fig:q2Tc}
\end{figure}

For three dimensional $2$-state Potts model, by feeding the processed data from Section \ref{dataset} and \ref{optimal}  into the DANN, the output are probabilities $P_0$ and $P_1$ for the configurations belonging to phase ``0" and phase ``1" respectively. 
For ease of presentation, we only analyze $P_0$ in the following, as $P_0+P_1=1$.
The average probability $P_0$ as a function of $T$ at $L=12$ is shown Fig.~\ref{Fig.q2.1}, along with a sigmoid function fitting(red dashed line)
specifically expressed as follows (1-sigmoid):
\begin{equation}
 T \to 1 - \frac1{1+e^{\frac{-(T-T_c)}{\sigma}}} \,,
 \label{eq:sigmoid}
\end{equation}
where $T_c$ is the critical temperature estimated by DANN and $\sigma$ is the scaling of the width. The shaded region in Fig.~\ref{Fig.q2.1} represents the optimal target domain, and the detailed process of obtaining the optimal source domain (or target domain) is descibed in Fig~\ref{Fig.q2.2} as the iterative method in Section \ref{optimal}. The estimate critical temperature of DANN captured at $P_0=0.5$ is $T_c=2.2113$.

\begin{figure}[htbp]
\centering
\subfigure[]{
\label{Fig.q2.3}
\includegraphics[height=0.3\textwidth]{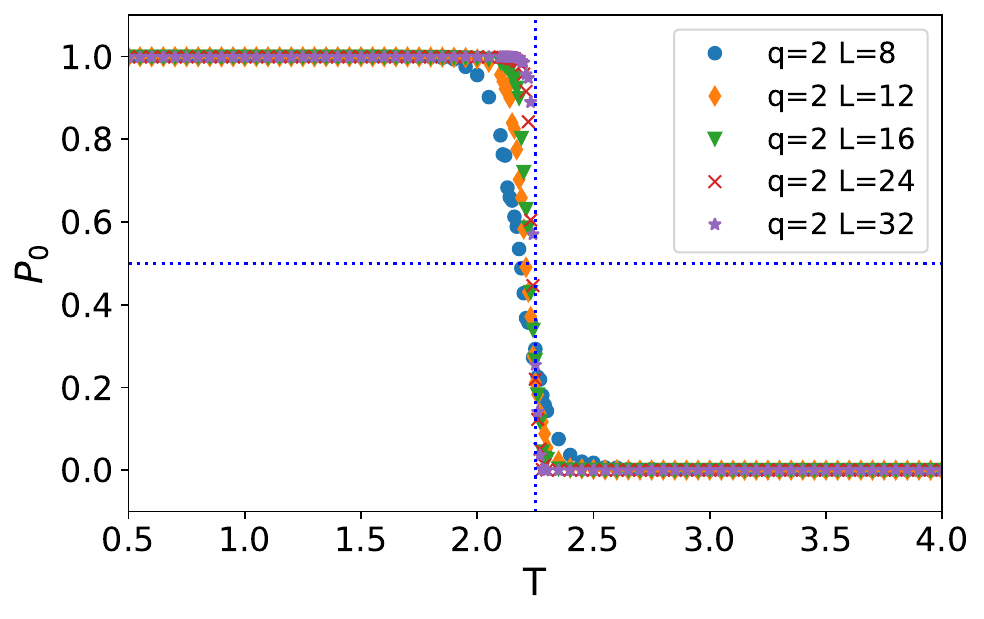}}
\subfigure[]{
\label{Fig.q2.4}
\includegraphics[height=0.3\textwidth]{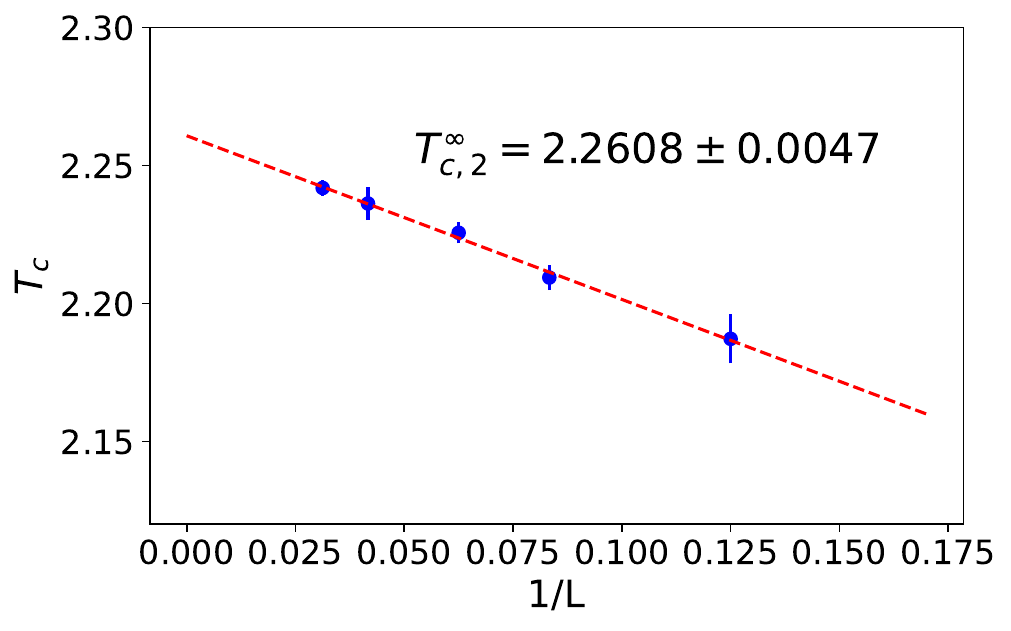}}
\caption{(a) Average probability $P_0$ of belonging to phase ``0" (ordered phase) for 3D $2$-state Potts model at $L = 8,12,16,24$ and $32$. (b) Extrapolation of the critical point $T_{c,q=2}$ to infinite lattice size. 
The errorbar is the standard deviation calculated from $5$ independent training sessions of the DANN.
}
\end{figure}

\begin{figure}[htbp]
\subfigure[]{
\label{Fig.q2.5}
\includegraphics[height=0.33\textwidth]{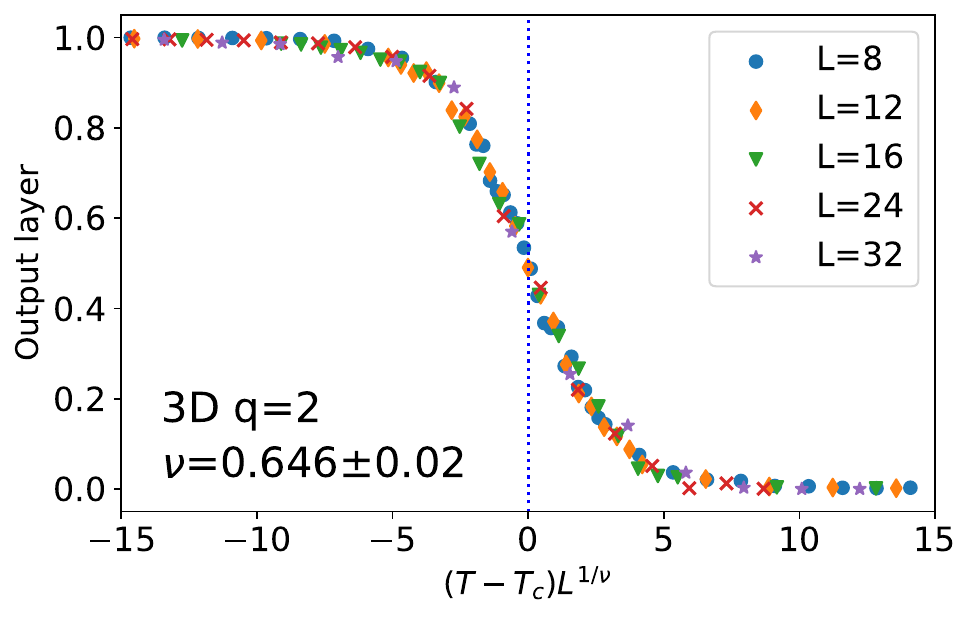}}
\subfigure[]{
\label{Fig.q2.6}
\includegraphics[height=0.33\textwidth]{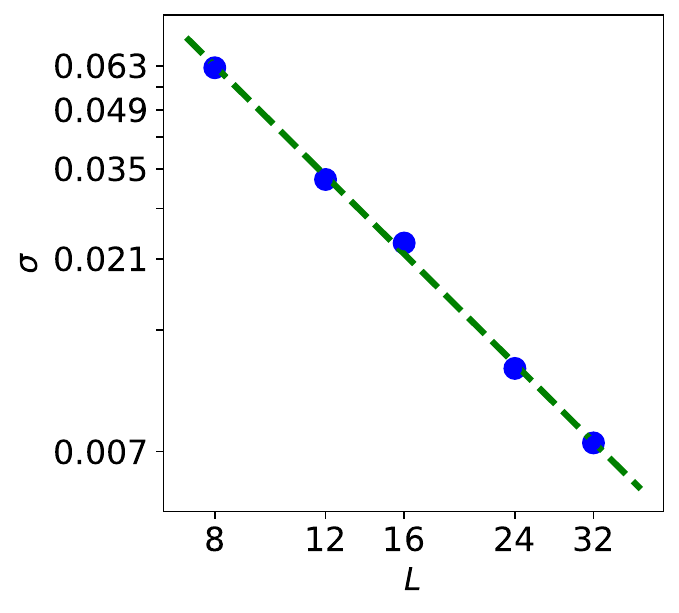}}
\caption{(a) Data collapse of output layer as a function of the scaling $(T-T_c)L^{1/ \nu}$ for 3D $2$-state Potts model with different lattice sizes. (b) The fitting of critical exponent $\nu$,  appearing as a straight line in the double logarithmic transformation of the sigmoid function parameter $\sigma$ (see Eq.~(\ref{eq:sigmoid})) and the system lattice size $L$.
}
\label{fig:q2v1}
\end{figure}

In Fig.~\ref{Fig.q2.3}, it shows that the method of DANN successfully classifies the low- and high-temperature phases of the $q=2$ state model at $L=8,12,16,24$ and $32$.
It is a continuous behavior of quantity $P_0$ in the area approaching $T_c$, which means the system of state $q=2$ is a second-order phase transition.
To extract the critical temperature of an infinite lattice system, 
we take the finite-size scaling (FSS) technique \cite{fisher1972scaling,barber1983finite,
privman1990finite,henkel1999conformal} to extrapolate the results of the different sizes in $L \in [8,32]$ to zero on the $1/L$ scale, as shown in Fig.~\ref{Fig.q2.4}, with a weighted linear fit.\footnote{Weighted linear regression with errors in $y$, a weighting factor: $w_i = \frac{\xi_i^{-2}}{\sum_i \xi_i^{-2}}$, where $\xi_i$ is the standard deviation for $y_i$.
}
The estimate critical temperature of infinite system is $T_{c,q=2}^{\infty}= 2.2608  \pm 0.0047 $, close to the MC result $T_c=2.2558$\cite{talapov1996magnetization}.

The critical exponent plays a key role in describing the critical behavior of phase transition.
We can use the data collapse method to obtain the critical exponent $\nu$. As shown in Fig.~\ref{Fig.q2.5}.
All data points of $P_0$ at $L=8,12,16,24$ and $32$ are collapsed to a single curve as a function of $(T-T_c)L^{1/\nu}$.
According to the sigmoid fitting of Eq.~\ref{eq:sigmoid} for $(T-T_c)L^{1/\nu}=(T-T_c)/\sigma$ as $\sigma$ chosen in Fig.~\ref{Fig.q2.6}, we can obtain the estimate $\nu \simeq 0.646 \pm 0.02$ by log-log transformation, consistent with the MC simulations of $ 0.630$ \cite{xu201892}.

\subsection{The DANN results of q=3}

\begin{figure}[htbp]
\centering
\subfigure[]{
\label{Fig.q3.3}
\includegraphics[height=0.3\textwidth]{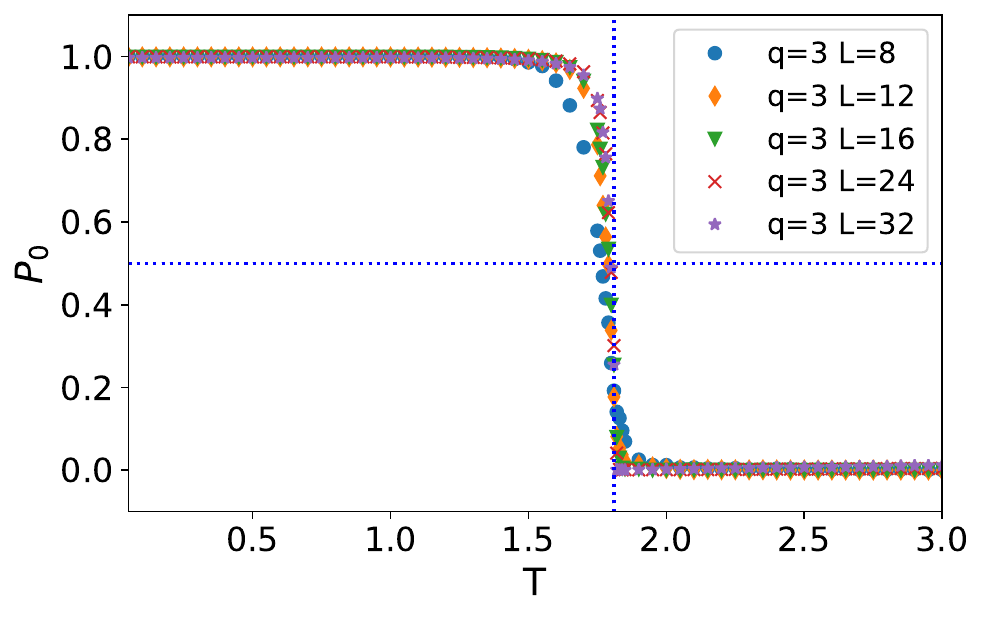}}
\subfigure[]{
\label{Fig.q3.4}
\includegraphics[height=0.3\textwidth]{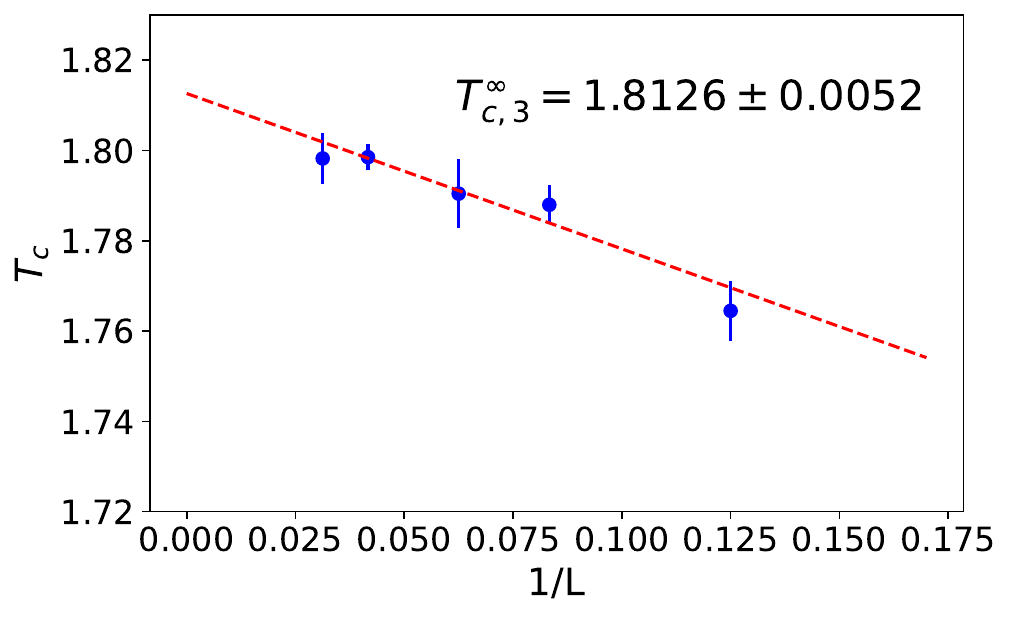}}
\caption{(a) Average probability of belonging to phase ``0" for two-dimensional $3$-state Potts model at $L = 8,12,16,24$ and $32$. (b) Extrapolation of the critical temperature $T_{c,q=3}$ to infinite lattice size.}
\label{fig:q3TC}
\end{figure}

By the successful experience of $q=2$, we apply the DANN to search $T_c$ in the case of $q=3$ with the size $L=8,12,16,24$ and $32$, as shown in Fig.~\ref{fig:q3TC}.
The estimated values of $T_{c,q=3}^{\infty}$ is $ 1.8126  \pm 0.0052 $, which is close to the results from traditional method $T_c=1.8163$ of Ref.~\cite{bazavov2007normalized} and  $T_c=1.8195$ of Ref.~\cite{gendiar2002latent}. From the DANN result $P_0$ in Fig~\ref{Fig.q3.3},  the phase transition of $q=3$ seems to be continuous observed through the eyes. However, it has been proved that the  type of phase transition for the cases of 
$q \geq 3$ belongs to first-order phase transition~\cite{hartmann2005calculation,lee1991three}. 
Obviously, the intuitive judgment is not a sufficient,
and a more convinced analysis for determining the phase type will be given in Section~\ref{sec:det_trans}.

\subsection{The DANN results of q=4 and 5}

\begin{figure}[htbp]
\centering
\subfigure[]{
\label{Fig.q4.1}
\includegraphics[height=0.29\textwidth]{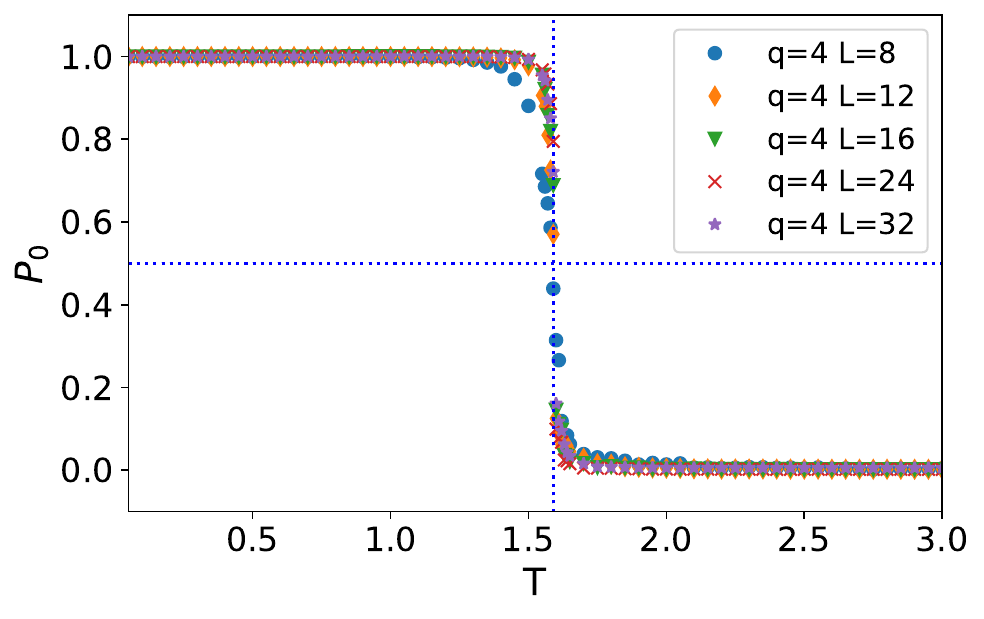}}
\subfigure[]{
\label{Fig.q4.2}
\includegraphics[height=0.29\textwidth]{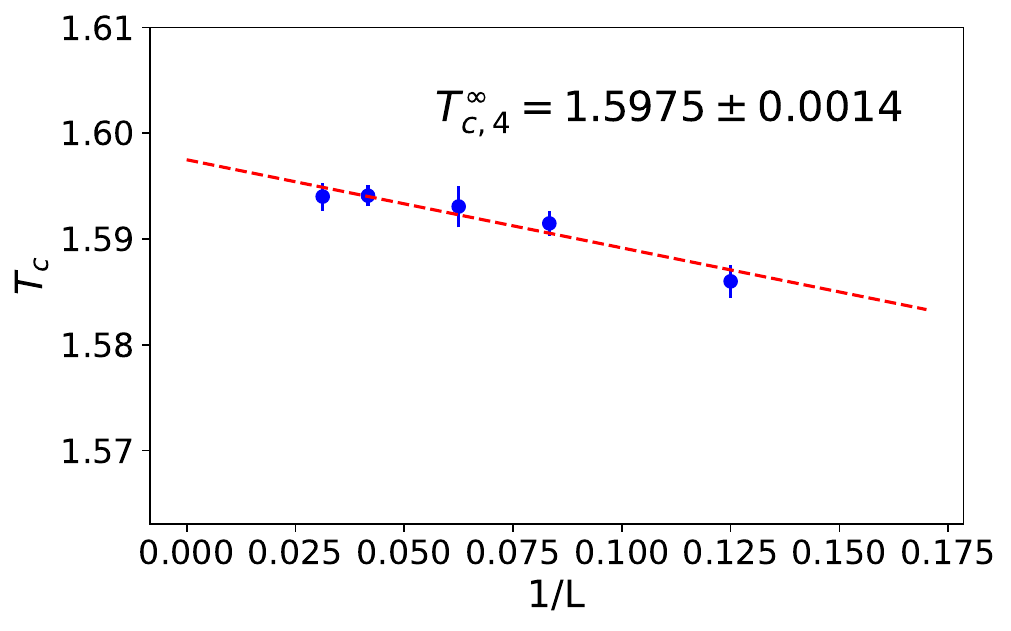}}
\subfigure[]{
\label{Fig.q5.1}
\includegraphics[height=0.29\textwidth]{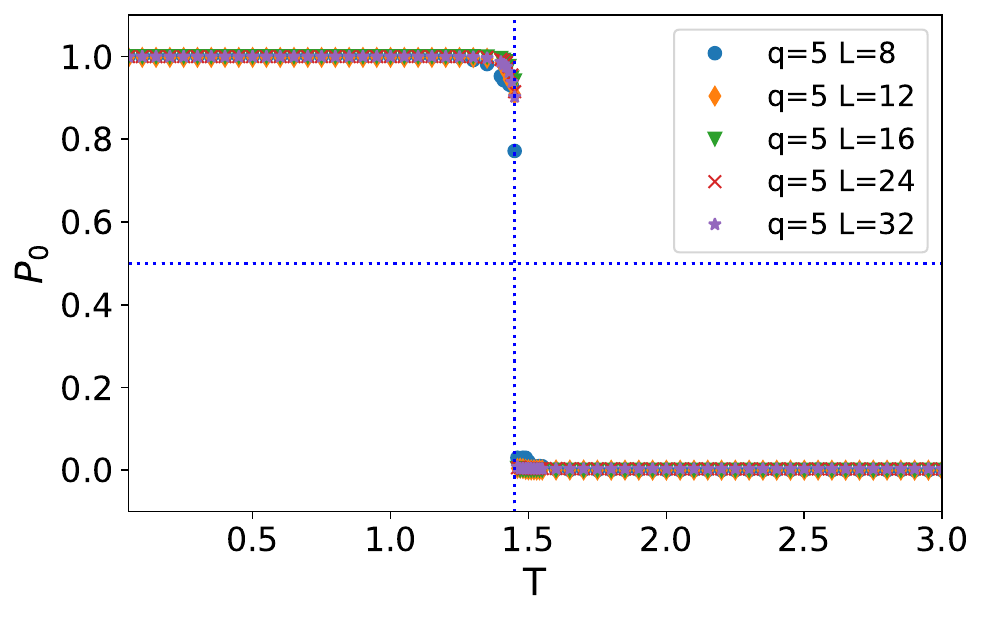}}
\subfigure[]{
\label{Fig.q5.2}
\includegraphics[height=0.29\textwidth]{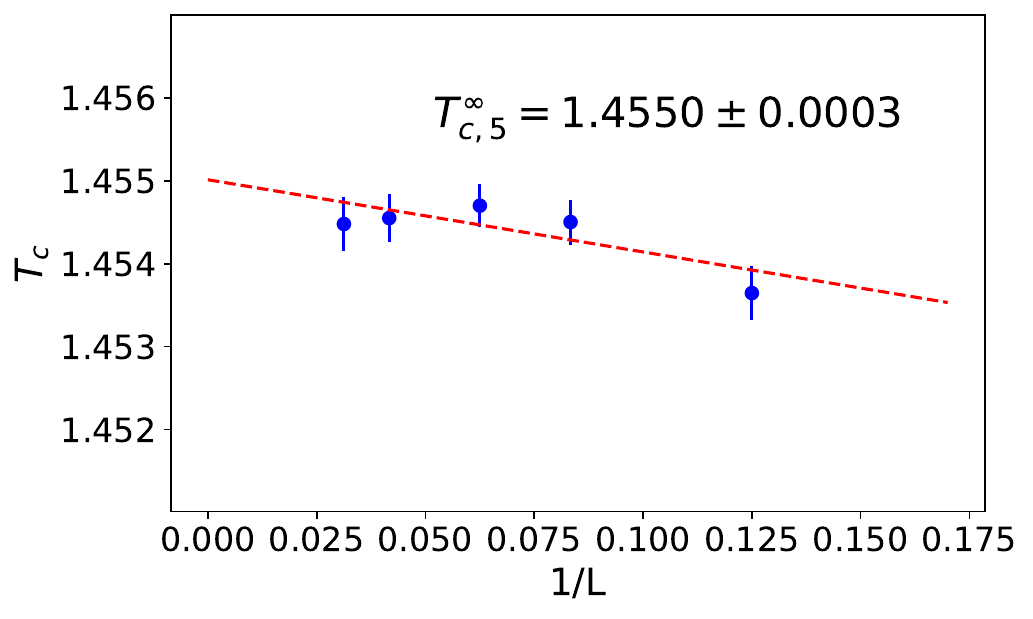}}

\caption{The average probability $P_0$ of three-dimensional Potts model for (a) $q= 4$, (c) $q=5$, with their extrapolation of the critical temperature to infinite lattice size (b) $T_{c,q=4}^\infty$, (d) $T_{c,q=5}^\infty$ by using FSS.
}
\label{fig:q4q5}
\end{figure}

Similarly, we apply DANN and conduct the tests for three dimensional $q$-state Potts model with $q=4$ and $5$, respectively. The relevant results of $P_0$ and FFS fitting are shown in Fig.~\ref{fig:q4q5}. In Fig.~\ref{Fig.q4.1} and Fig.~\ref{Fig.q5.1}, we can see 
the distribution of the average probability $P_0$ becomes significantly discontinuous in the area close to $T_c$, indicating that  
the system directly transition from an ordered phase to an disordered phase.
This jumping phenomenon implies the existence of a first-order phase transition, which is consistent with the MC predictions as the cases of $q=4$ and $5$.
And the predicted critical temperatures $T_{c,4}^\infty= 1.5975 \pm 0.0014$ and $T_{c,5}^\infty= 1.4550 \pm 0.0003$ shown in Fig.~\ref{Fig.q4.2} and Fig.~\ref{Fig.q5.2} are also similar to the results obtained by MC method.
All of the DANN results of $q=2,3,4$ and $5$ are listed in Table.~\ref{tab:NumericalResults}, compared with the results of MC and supervised learning.

\begin{table}[htbp]
\caption{Critical properties of the 3D ferromagnetic $q$-state Potts model. $T_{c,q}$ is the value obtained from Monte Carlo method. The values of $T^{'}_{c,q}$ are the results of supervised learning.  $T_{c,q}^\infty$ is the critical temperature predicted by DANN for infinite systems, and the transition types obtained using DANN trained with spin configuration samples, are second-order (2nd) and first-order (1st). } 
\label{tab:NumericalResults}
\centering
\begin{tabular}{lcccccccc}
\hline\hline
q & $T_{c,q}$ \cite{bazavov2008phase,talapov1996magnetization,bazavov2007normalized}
& $T^{'}_{c,q}$ \cite{yau2022generalizability} & $T_{c,q}^\infty$   &transition type&\\ \hline

2 & 2.2558& 2.2502(31) & 2.2608(47) &2nd&\\
3 & 1.8163& 1.8188(107)& 1.8126(52) &1st&\\
4 & 1.5908& 1.5903(6)& 1.5975(14)& 1st & \\
5 & 1.4504& ... & 1.4550(3)&1st&\\ 

\hline\hline
\end{tabular}
\end{table}

\subsection{The order of the phase transition}
\label{sec:det_trans}
 
In previous sections, we provide an accurate estimates of $T^{\infty}_c$ and critical exponent $\nu$ of the 3D $q$-state Potts model via the DANN, and also
used an intuitive way to determine the type of phase transition.

However, the sudden jump of the observables near critical temperatures is not a sufficient evidence to judge the first- or second-order phase transition, and a convinced method has been approved: checking the existence of the metastable behaviors~\cite{chen2023study,d2022phase}. In detail, for a first order transition, a two-peaked distribution of the $P_0$'s of individual configurations happens near $T_c$, and no obvious two-peaked structure for the cases of a second order phase transition.

\begin{figure}[htbp]
\centering
\subfigure[$q=2,T=2.16$]{
\label{Fig.disPq2.1}
\includegraphics[height=0.23\textwidth]{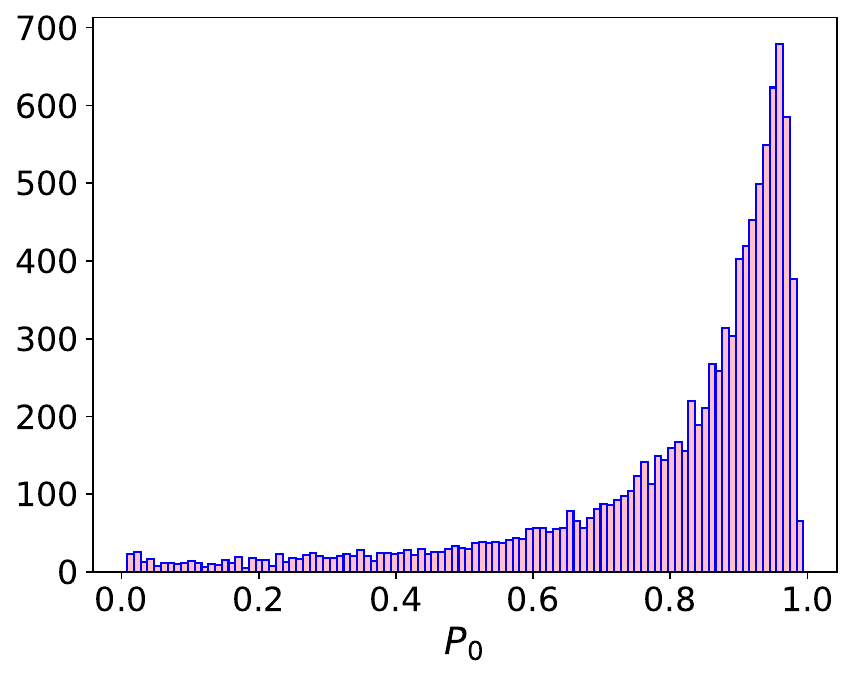}}
\subfigure[$q=2,T=2.2$]{
\label{Fig.disPq2.2}
\includegraphics[height=0.23\textwidth]{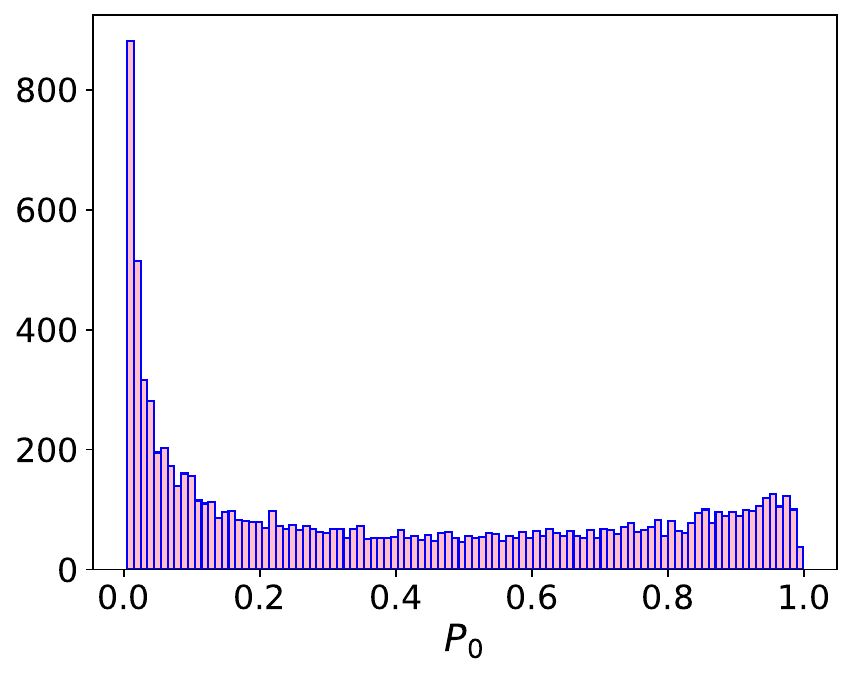}}
\subfigure[$q=3,T=1.76$]{
\label{Fig.disPq3.1}
\includegraphics[height=0.23\textwidth]{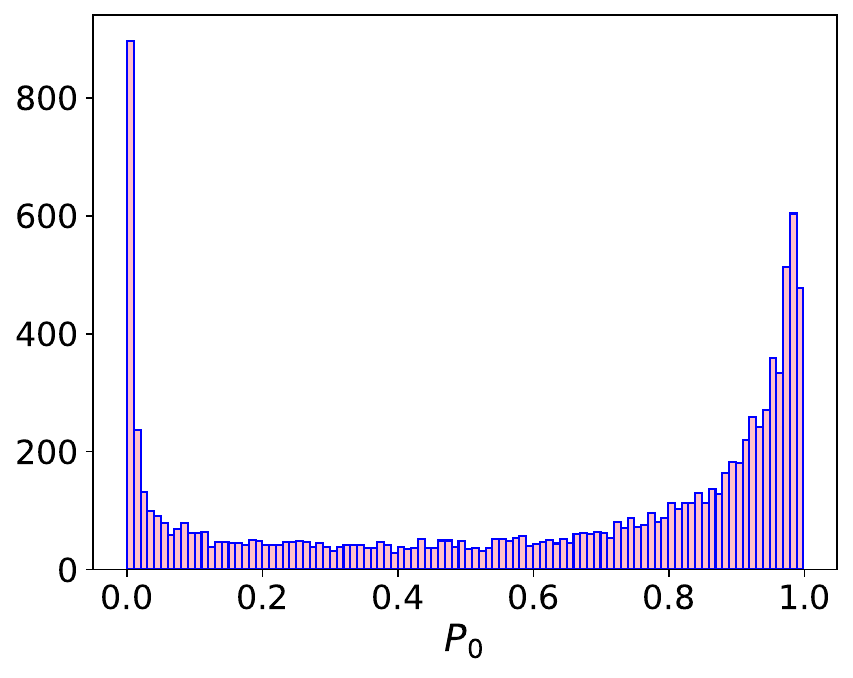}}
\subfigure[$q=3,T=1.77$]{
\label{Fig.disPq3.2}
\includegraphics[height=0.23\textwidth]{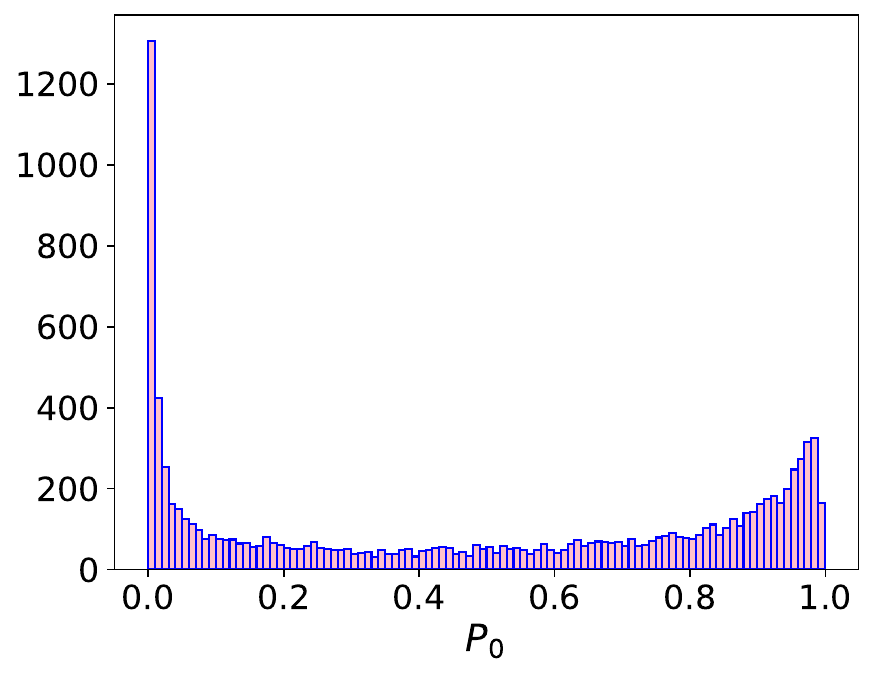}}
\subfigure[$q=4,T=1.585$]{
\label{Fig.disPq3.1}
\includegraphics[height=0.23\textwidth]{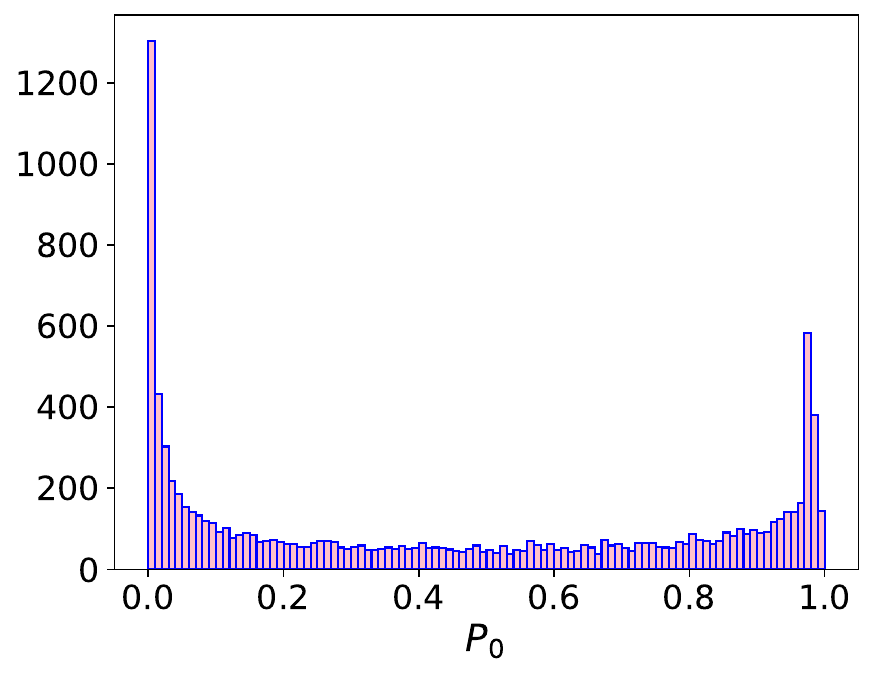}}
\subfigure[$q=4,T=1.59$]{
\label{Fig.disPq4.2}
\includegraphics[height=0.23\textwidth]{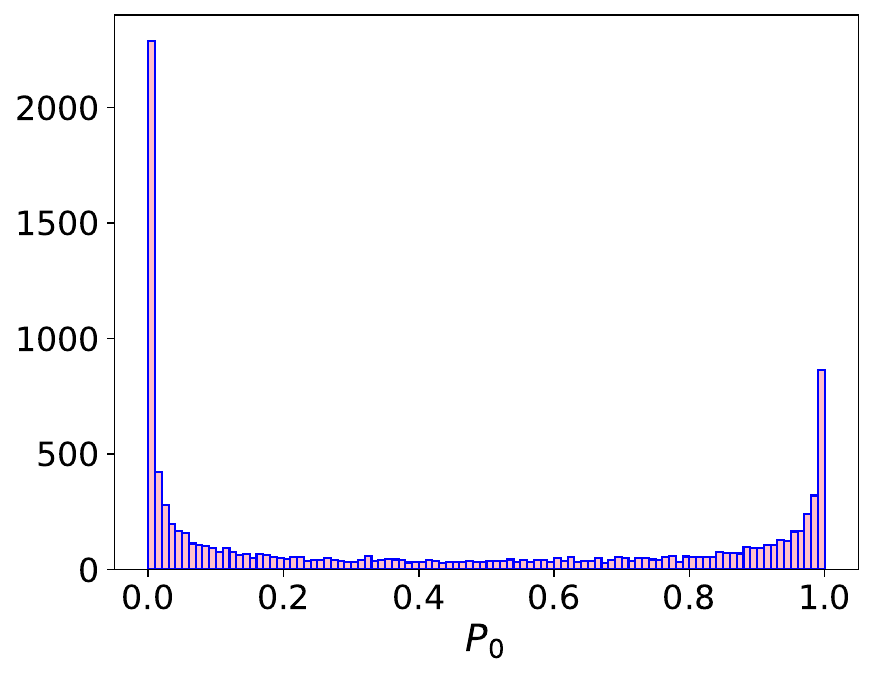}}
\subfigure[$q=5,T=1.4535$]{
\label{Fig.disPq5.1}
\includegraphics[height=0.23\textwidth]{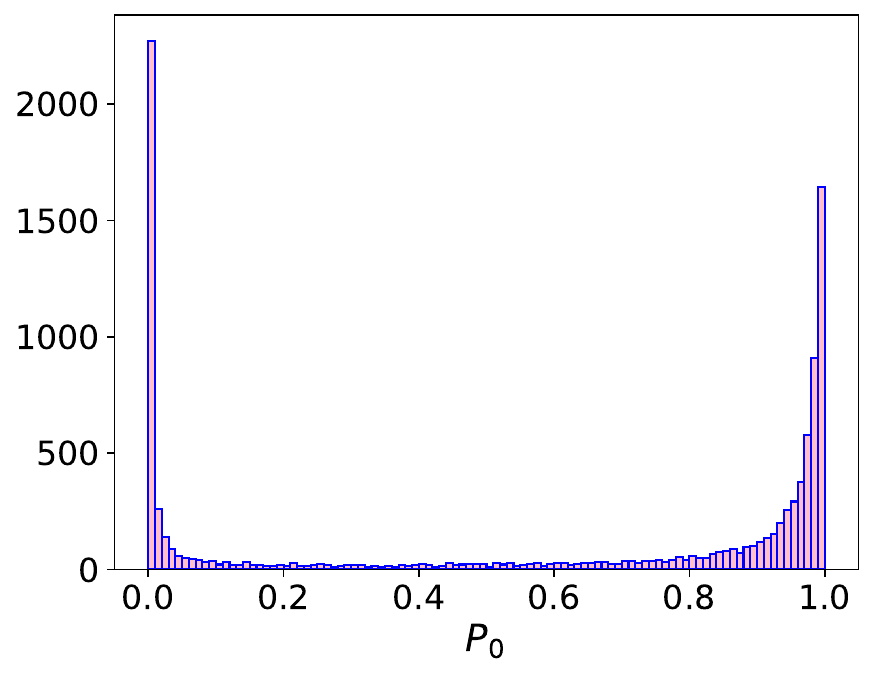}}
\subfigure[$q=5,T=1.454$]{
\label{Fig.disPq5.1}
\includegraphics[height=0.23\textwidth]{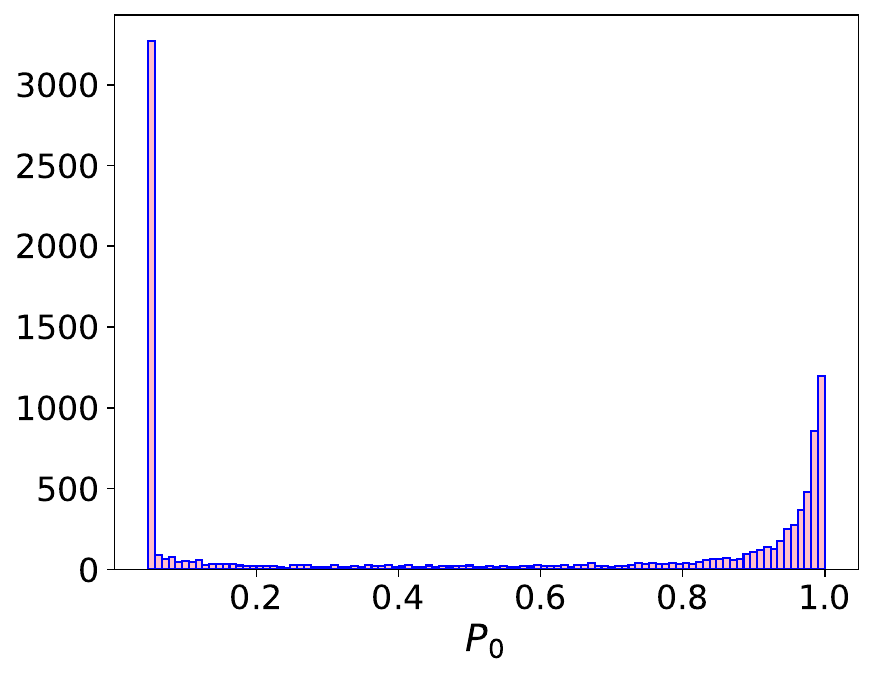}}
\caption{The histograms of DANN output $P_0 $ for each state with size $L=8$. $10000$ data points are generated for each of the plots, and each data point represents the output $P_0$ of a single configuration sample evaluated by DANN. }
\label{fig:DANNPdisL8}
\end{figure}
\begin{figure}[htbp]
\centering
\subfigure[$q=2,T=2.20$]{
\label{Fig.l12disPq2.1}
\includegraphics[height=0.23\textwidth]{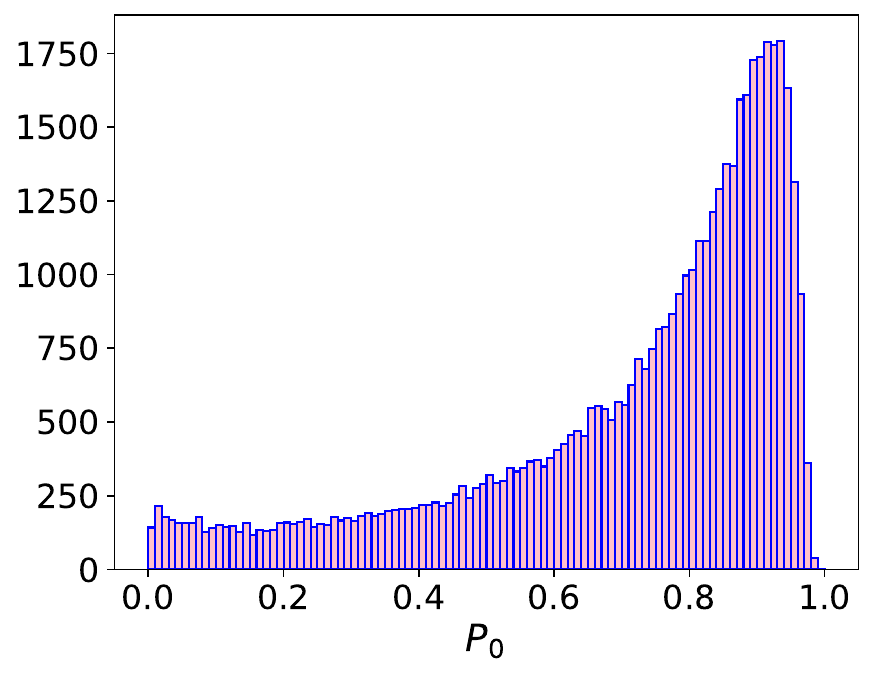}}
\subfigure[$q=2,T=2.215$]{
\label{Fig.l12disPq2.2}
\includegraphics[height=0.23\textwidth]{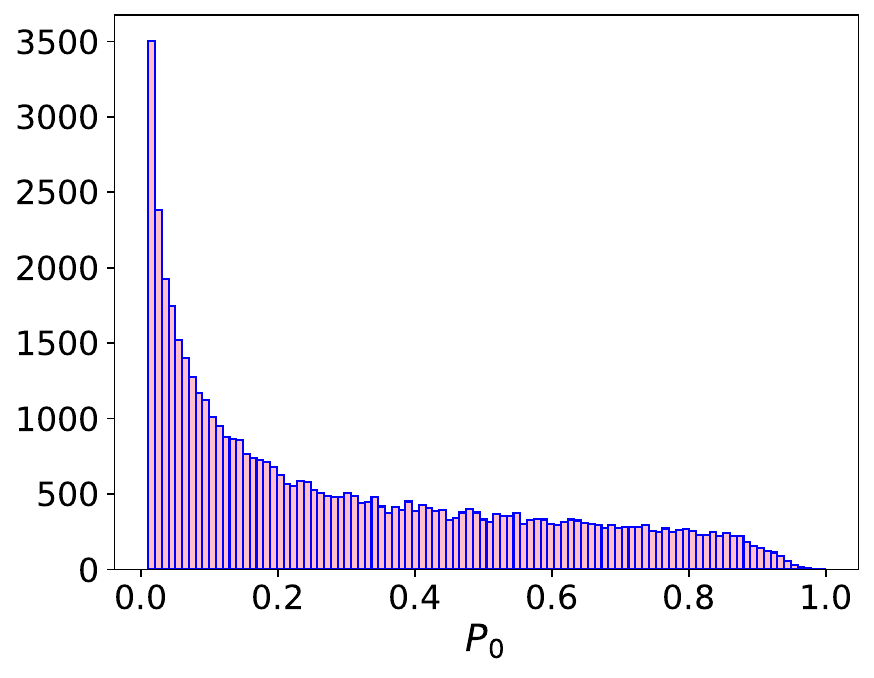}}
\subfigure[$q=3,T=1.785$]{
\label{Fig.l12disPq3.1}
\includegraphics[height=0.23\textwidth]{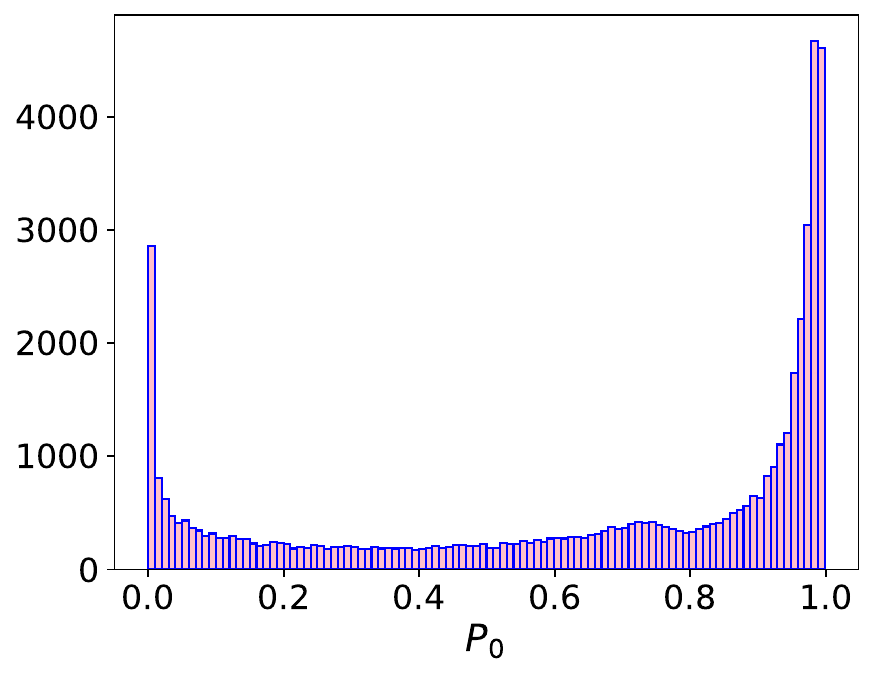}}
\subfigure[$q=3,T=1.79$]{
\label{Fig.l12disPq3.2}
\includegraphics[height=0.23\textwidth]{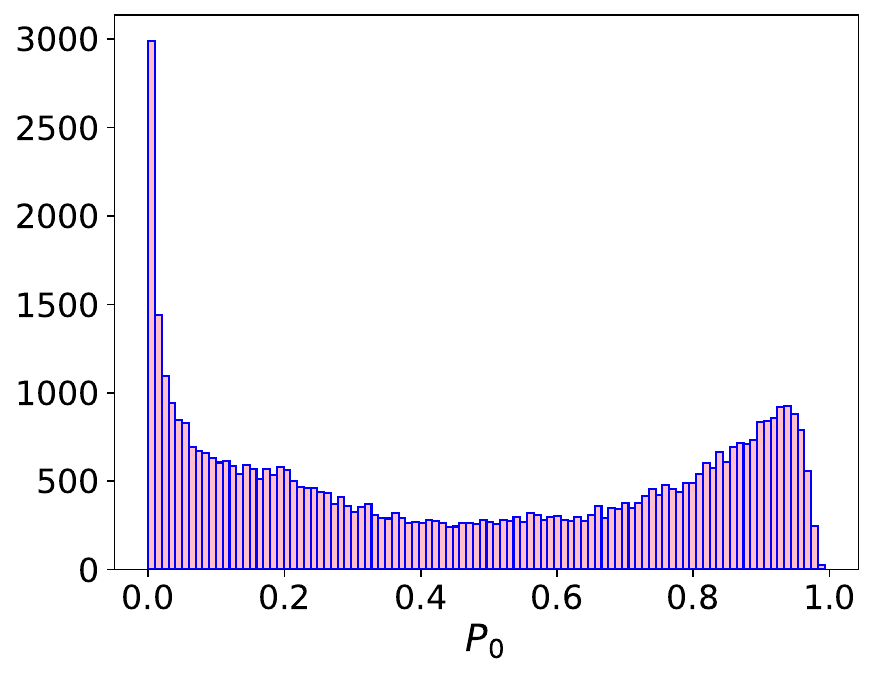}}
\subfigure[$q=4,T=1.591$]{
\label{Fig.l12disPq3.1}
\includegraphics[height=0.23\textwidth]{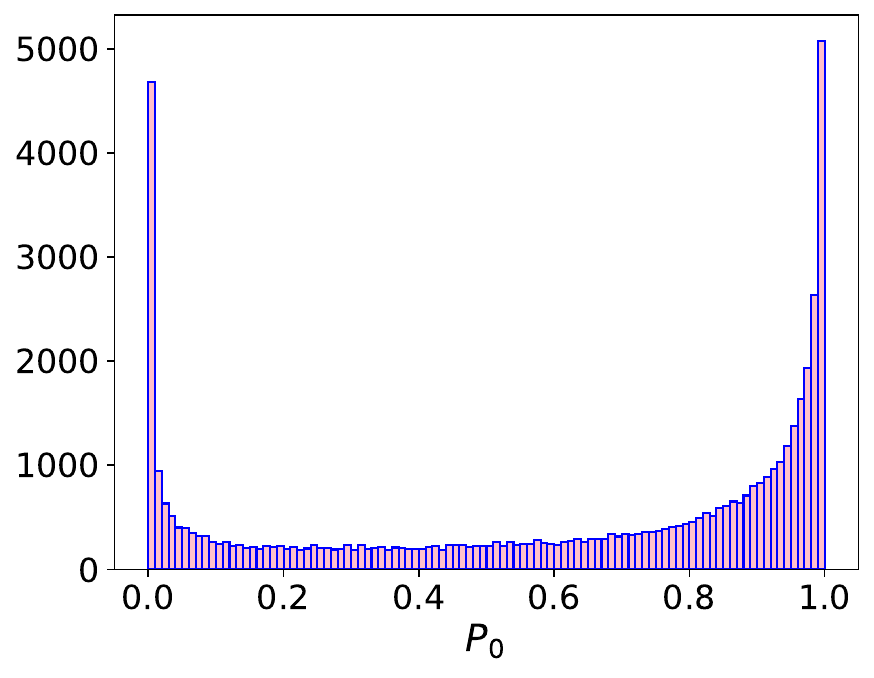}}
\subfigure[$q=4,T=1.592$]{
\label{Fig.l12disPq4.2}
\includegraphics[height=0.23\textwidth]{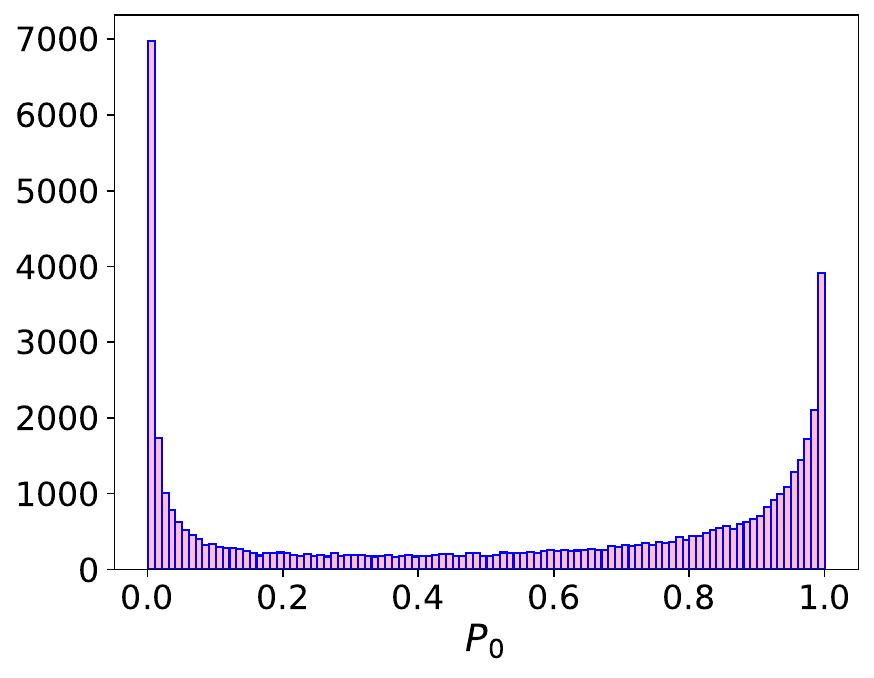}}
\subfigure[$q=5,T=1.4545$]{
\label{Fig.l12disPq5.1}
\includegraphics[height=0.23\textwidth]{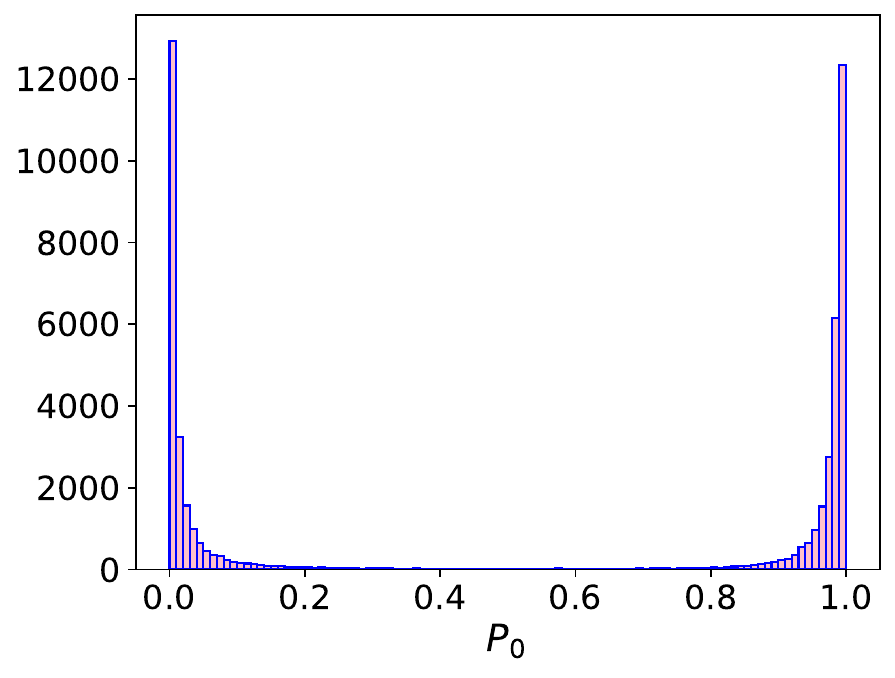}}
\subfigure[$q=5,T=1.455$]{
\label{Fig.l12disPq5.1}
\includegraphics[height=0.23\textwidth]{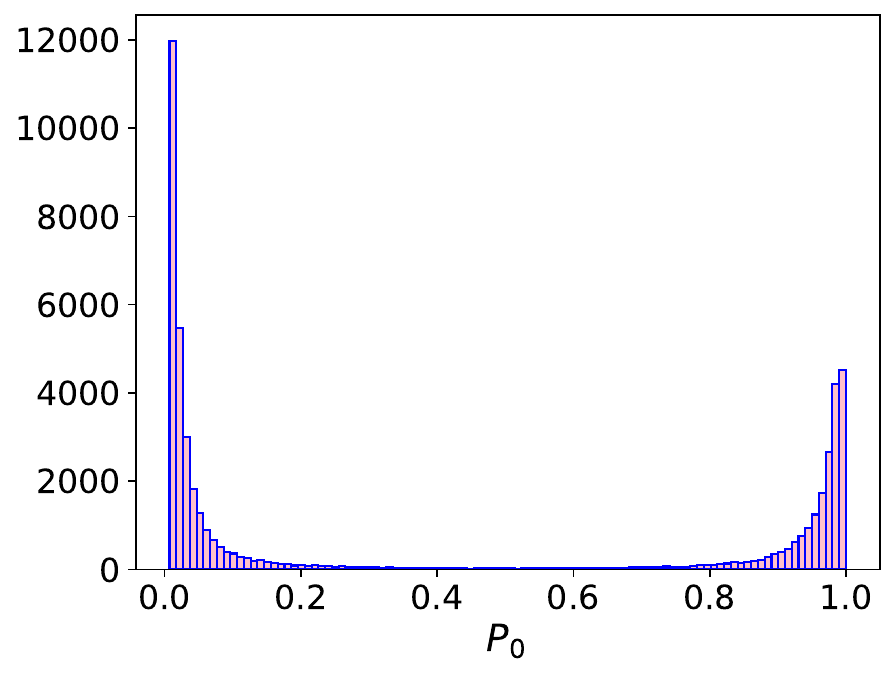}}
\caption{The histograms of DANN output $P_0 $ for each state with size $L=12$. $50000$ data points are generated for each of the plots, and each data point represents the output $P_0$ of a single configuration sample evaluated by DANN. }
\label{fig:DANNPdisL12}
\end{figure}

To demonstrate this approach, $10000$ spin configurations at two temperatures close to the estimated $T_{c,q}$ are generated for each state ($q=2,3,4$ and $5$) with $L=8$. 
From the the histogram of DANN output $P_0$ at the corresponding temperature as demonstrated in Fig.~\ref{fig:DANNPdisL8}.
It can be seen that the distribution of $P_0$ is a structure of single peak for $q=2$, while the two-peaked distribution appear for $q=3,4,5$, which clearly indicate that the former belongs to second order phase transition, and the latter are the cases of first order phase transition. To ensure that the results are not affected by size $L$, we also show the histograms for $L=12$ in Fig.~\ref{fig:DANNPdisL12}, which is consistent with $L=8$.

\section{Conclusion}

We have employed MC simulations and a semi-supervised machine learning method to study the phase transition characteristics in the 3D $q$-state ferromagnetic Potts model on a cubic lattice.
The states considered here are $q=2,3,4,5$.
The MC calculations evaluated the phase transition types of each state as expected. It is worth noting that,
our ML method, DANN, is perfectly suitable for studying the related phase transitions for different state models. 

DANN has a unique deep learning structure, consisting of a feature extractor $G_f$, a label predictor $G_y$, and a domain classifier $G_d$, which makes it capable of adversarial learning.
Therefore, in our work, unlike other supervised learning methods that rely on fully labeled samples,  
we utilize only a few automatically labeled configuration samples, leading to a significant reduction in computational cost.
Moreover, DANN does not require any prior information about the system during the training process. Instead, it can locate the critical temperature based solely on the raw spin configuration samples.

During the operation of DANN, we employ the iterative method to derive the optimal source and target domains. This approach enables DANN to efficiently and precisely learn the most relevant information, and achieve optimal results.
Finally, it was found that, DANN demonstrated remarkable success in distinguishing between the high-temperature phase (ordered phase) and low-temperature phase (disordered phase) for each state model.

Notably, in the case of $q=2$, DANN not only accurately localized the critical temperature $T_c$ but also obtained the critical exponent $\nu$ through the data collapse method (namely, $T_{c,2}^\infty= 2.2608 \pm 0.0047$, $\nu=0.646 \pm 0.02$). 
At the same time, the critical temperature of $q=3,4$ and $q=5$ is also successfully identified (namely, $T_{c,3}^\infty= 1.8126 \pm 0.0052$, $T_{c,4}^\infty= 1.5975 \pm 0.0014$, and $T_{c,5}^\infty= 1.4550 \pm 0.0003$ respectively).
All of which were found to be in agreement with the Monte Carlo results.

In addition, it was found that using the same network structure, DANN can also clearly determine the types of phase transitions in different states: first-order or second-order phase transition. By studying the distribution of the $P_0$’s (DANN outputs) of individual configurations around the predicted critical temperature, we found that, for a second order phase transition of $q=2$, the histogram of $P_0$ has a single peak structure, but the two-peaked distributions
appear for $q = 3, 4$ and $5$ of the first order phase transitions. 
This implies that the network successfully captured potential unique information during the training process: not only extracting information about the second-order phase transition, but also identifying the coexistence of two phases in the first-order phase transition.

The results obtained on the 3D Potts model once again demonstrate the power and stability of DANN.
At present, for unknown systems or other systems where accurate order parameters cannot be obtained, it becomes extremely difficult to study the phase transition using traditional methods. Therefore, we hope that DANN can serve as an alternative method to explore new models or fields, providing a reference value for subsequent theoretical and numerical methods.
Even utilizing its adversarial learning ability to extract information and transfer features to high-dimensional fields may achieve unexpected results, which is also one of our future research topics.

\section*{Acknowledgements}
We gratefully acknowledge the fruitful discussions with Shengfeng Deng, Dian Xu and Kui Tuo. 
This work was supported in part by National Natural Science Foundation of China (Grant No. 61873104, 11505071), the Programme of Introducing Talents of Discipline to Universities under Grant no. B08033, the Fundamental Research Funds for the Central Universities, and the European Union project 
RRF-2.3.1-21-2022-00004 within the framework of MILAB.

\section*{Author contributions statement}
Xiangna Chen: Investigation, Conceptualization, Methodology, Validation, Writing - Original Draft.
Feiyi Liu: Methodology, Software, Writing - Review and Editing, Supervision. 
Weibing Deng: Conceptualization, Writing - Review and Editing, Supervision, Funding acquisition.
Shiyang Chen: Software, Writing - Review and Editing. 
Jianmin Shen: Conceptualization, Methodology.
Gabor Papp: Software, Validation, Writing - Review and Editing, Funding acquisition.
WeiLi: Conceptualization, Resources, Supervision.
Chunbin Yang: Conceptualization, Supervision.
All authors reviewed the manuscript. 



\bibliography{DANN3d}

\end{document}